\documentclass{aa}  

\usepackage{graphicx}
\usepackage{txfonts}

\usepackage{siunitx}
\usepackage[utf8]{inputenc} % usually not needed (loaded by default)
\usepackage[T1]{fontenc}
\usepackage{natbib}
\usepackage{bm}
\usepackage{dsfont}
\usepackage{amsmath}
\usepackage{graphicx}	% Including figure files
\usepackage{txfonts}
\usepackage{subfigure}
\usepackage{ulem}
\newcommand{\res}[1]{{#1}}
\newcommand{\response}[1]{{#1}}

\newcommand*{\occr}{\ensuremath{\eta}}
\newcommand*{\tr}{\ensuremath{\eta _{tr}}}
\newcommand*{\ds}{\ensuremath{\eta _{det}}}
\newcommand*{\comp}{\ensuremath{H}}
\newcommand*{\snrlim}{7}

\DeclareSIUnit[]\Mj
 {\text{\ensuremath{M_{\textup{J}}}}}
\DeclareSIUnit[]\Rj
 {\text{\ensuremath{R_{\textup{J}}}}}
\DeclareSIUnit[]\Re
 {\text{\ensuremath{R_{\oplus} } } }
\DeclareSIUnit[]\Me
 {\text{\ensuremath{M_{\oplus} } } }
\DeclareSIUnit[]\Rsun
 {\text{\ensuremath{R_{\odot} } } }
\DeclareSIUnit[]\Msun
 {\text{\ensuremath{M_{\odot} } } }
\DeclareSIUnit[]\Lsun
 {\text{\ensuremath{L_{\odot} } } }

\begin{document}

   \title{The occurrence rate of exoplanets orbiting ultracool dwarfs as probed by \textit{K2}}

   % \subtitle{Ultracool dwarf planets}

   \author{Marko Sestovic
          \inst{1}
          \and
          Brice-Olivier Demory\inst{1}\fnmsep
          %\thanks{Just to show the usage of the elements in the author field}
          }

   \institute{Centre for Space and Habitability, University of Bern,
              Gesellschaftsstrasse 6, Bern, Switzerland\\
              \email{marko.sestovic@csh.unibe.ch}
             }

   \date{Received 14 February 2020; accepted 29 April 2020}

% \abstract{}{}{}{}{} 
% 5 {} token are mandatory
 
  \abstract
  % context heading (optional)
  % {} leave it empty if necessary  
   {With the discovery of a planetary system around the ultracool dwarf TRAPPIST-1, there has been a surge of interest in such stars as potential planet hosts. Planetary systems around ultracool dwarfs represent our best chance of characterising temperate rocky-planet atmospheres with JWST.
   %\textbf{Previous studies have also shown that short-period small planets become more common as stellar mass decreases.}
   However, TRAPPIST-1 remains the only known system of its kind, and the occurrence rate of planets around ultracool dwarfs is still poorly constrained.}
  % aims heading (mandatory)
   {We seek to perform a complete transit search on the ultracool dwarfs observed by NASA's K2 mission, and use the results to constrain the occurrence rate of planets around these stars.}
  % methods heading (mandatory)
   {We filter and characterise the sample of ultracool dwarfs observed by \textit{K2}, by fitting their spectral energy distributions, and using parallaxes from \textit{Gaia}. We build an automatic pipeline to perform photometry, detrend the lightcurves, and search for transit signals. Using extensive injection-recovery tests of our pipeline, we compute the detection sensitivity of our search, and thus the completeness of our sample. We infer the planetary occurrence rates within a hierarchical Bayesian model (HBM) to treat uncertain planetary parameters. With the occurrence rate parametrised by a step-wise function, we show a convenient way to directly marginalise over the second level of our HBM (the planetary parameters). Our method is applicable generally and can greatly speed up inference with larger catalogues of detected planets.}
  % results heading (mandatory)
   {We detect one planet in our sample of \res{702} ultracool dwarfs: a previously-validated mini-Neptune. We thus infer a mini-Neptune ($2-4\si{\Re}$) occurrence rate of \res{$\occr = 0.20^{+0.16}_{-0.11}$} within orbital periods of $1-20$ days. For super-Earths ($1-2\si{\Re}$) and ice/gas giants ($4-6\si{\Re}$) within $1-20$ days, we place \response{95\% credible intervals of \res{$\occr < 1.14$} and \res{$\occr < 0.29$}}, respectively. If TRAPPIST-1-like systems were ubiquitous, we would have had a \res{$\sim96\%$} chance of finding at least one.}
  % conclusions heading (optional), leave it empty if necessary 
   {}

   
   \titlerunning{Ultracool dwarf planets}
   \authorrunning{M. Sestovic \& B.-O. Demory}

   \maketitle
%
%-------------------------------------------------------------------

\section{Introduction} \label{sec:intro}

% TODO: Possibly introduce method - we use an HBM to accurately treat uncertainties in planet parameters, and we find the detection sensitivity of our full pipeline downstream of the photometry. "We have conducted an *empirical study of the detection efficiency of our search* through injection-recovery modelling." See also the wording used in DFM.
% TODO: Do I say more about why this is important for formation? Perhaps look at the Mulders papers and pull out the real points where the trend in occurrence rates could enlighten formation theory. e.g Mini-Neptunes are interesting because they contain 95% of the heavy element mass of M Dwarf planetary systems (Mulders 2015, increase in mass...)
%

The recent discovery of a seven-planet system around the M8V dwarf TRAPPIST-1 has attracted significant interest from the exoplanet community \citep{gillon2016,gillon2017}. These planets, all rocky and roughly the size of the Earth, form the longest chain of Laplace resonances currently known \citep{luger2017,grimm2018}. But perhaps its most important feature is that TRAPPIST-1 currently represents our best opportunity in the near future to characterise the atmospheres of Earth-sized planets in or near the habitable zone \citep{nutzman2008,seager2013,morley2017}.

Key to this are the properties of the star. TRAPPIST-1 belongs to the class of objects known as ultracool dwarfs: late-M and early-L dwarfs with radii comparable to Jupiter. As a result, Earth-sized planets occult on the order of 1\% of the ultra-cool star stellar disk when they transit. The transit depths of TRAPPIST-1 planets are therefore large enough that the upcoming James Webb Space Telescope may be able to detect the presence and composition of an atmosphere \citep[][]{seager2013,morley2017,lustig-yaeger2019,fauchez2020}.

\begin{figure}
    \centering
    \includegraphics[width=\hsize]{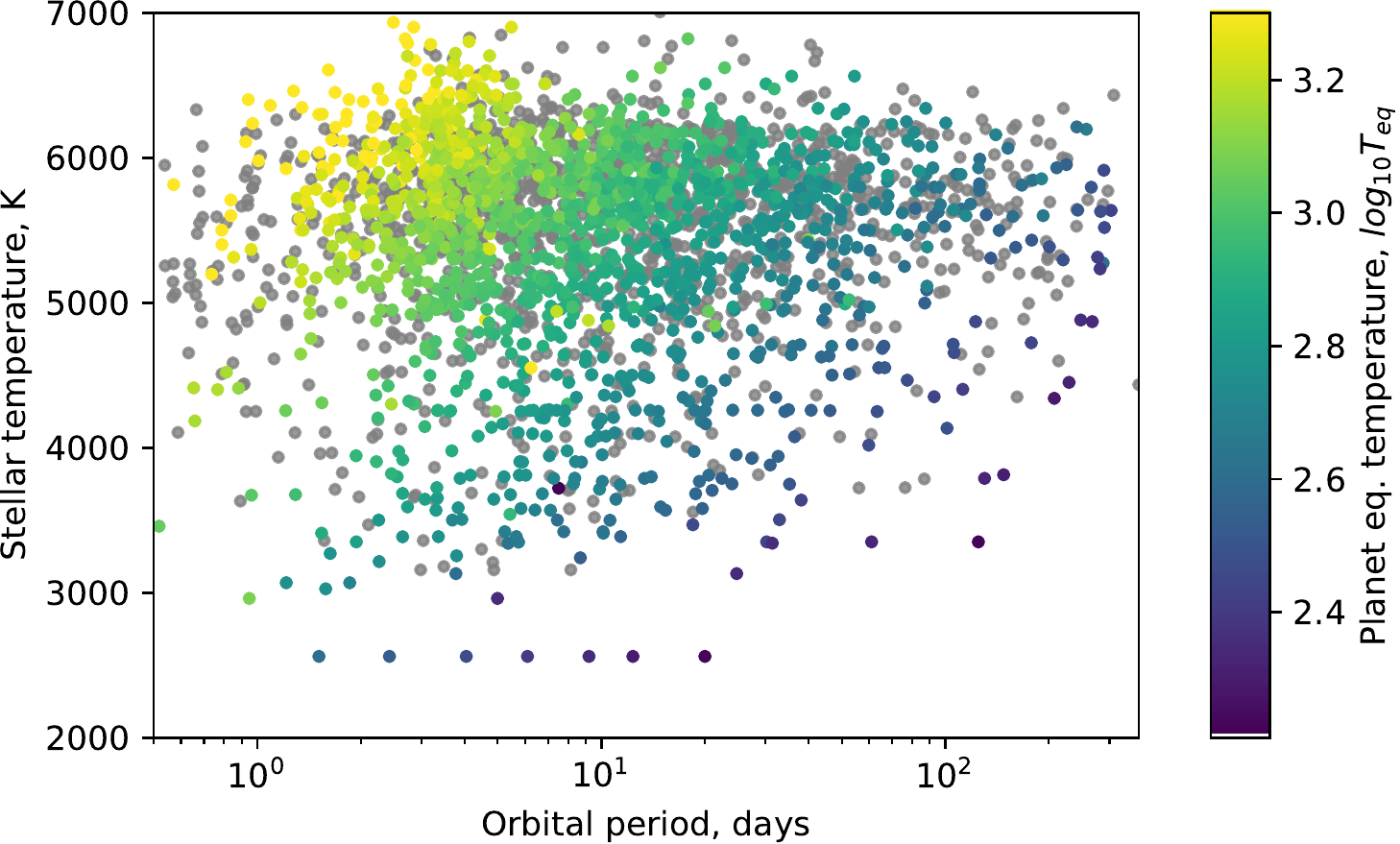}
        \caption{The catalogue of exoplanets discovered to-date, showing the diversity of stellar temperatures and orbital periods. In colour: the planet equilibrium temperature (grey for planets where it isn't given). The line of seven points at the bottom is the TRAPPIST-1 system. The catalog is taken from the confirmed planets discovered by RV and transits, from exoplanet.eu (updated 23/12/19).}
    \label{fig:intro-catalog}
\end{figure}

Nearly five years after its initial discovery, however, TRAPPIST-1 remains unique. There is a significant gap in the parameter space between the seven TRAPPIST-1 planets and the rest of the currently known M dwarf planets (see  Fig. \ref{fig:intro-catalog}). The detection of TRAPPIST-1 has sparked several projects to search for more such systems, such as SPECULOOS \citep{delrez2018}. It is timely to estimate the number of planets these surveys will detect, in order to inform their observational strategies. We therefore need to know the true planet occurrence rate, taking into account the strong observational biases impacting such surveys.

% From the planet formation perspective: at the low mass limit of the stellar main sequence, ultracool dwarfs provide a key test of trends previously found in the literature. In a review, \citet{mulders2018} show that small planets, with radii less than one Neptune radius, $\sim 4 \si{\Re}$, follow a trend of rising occurrence rates with decreasing stellar mass. For short periods ($\leq 50d$), the occurrence rate of such planets jumps by a factor of $\sim 3$ from the FGK stars to the M dwarfs \citep{mulders2018}. This trend has been explored into the early and mid-M dwarfs, taken from studies such as \citet{fressin2013,dressing2015} \tocite. Whether the trend continues into the ultracool dwarfs is still unknown. 

The advent of photometric and RV surveys such as \textit{Kepler}/\textit{K2} and \textit{HIRES} or \textit{HARPS} planet searches, among many others, has enabled the observations of thousands of stars and led to more than 3000 confirmed planet detections. With such a sample size, the exoplanet community has managed to precisely constrain the planet occurrence rates for stars ranging from F, G, K, down to early- and mid-M dwarfs; see e.g. \citet{youdin2011,howard2012,fressin2013,dressing2013,kopparapu2013,morton2014,dressing2015,ballard2016,mulders2018_EPOS,hsu2019}. Remarkably, in a review of these results, \citet{mulders2018} highlights the fact that occurrence rates of sub-Neptunes ($R < 4 \si{\Re}$) rise with decreasing stellar mass - by a factor of 3 from the FGK stars to the M dwarfs. This trend has been explored down into the early and mid-M dwarfs, by e.g \citet{berta2013,dressing2015,mulders2015a,muirhead2015,hardegree-ullman2019,hsu2020}. Ultracool dwarfs, at the lower-mass limit of the stellar main sequence, provide a key test of whether this trend continues. However, due to the ultracool dwarfs' intrinsic faintness, their planetary occurrence rates are more challenging to constrain.

Previously, \citet{demory2016} performed a search through 189 ultracool targets in early \textit{K2} Campaigns 1-6, finding no transiting planets. With an injection test, they showed that only 10\% of the targets in their sample exhibited a photometric precision amenable to the detection of a TRAPPIST-1b analogue. Using high-precision \textit{Spitzer} photometry, \citet{he2017} placed upper limits on the occurrence rate of planets around brown dwarfs, finding that there are fewer than $0.67$ planet per star, with radii between 0.75 and 3.25 \si{\Re} and periods of less than 1.28 days. However, as \textit{Spitzer} only observes one target at a time, the study was limited to short periods and a sample size of 44 targets. \response{During the writing of this publication, we were made aware of a work by \citet{sagear2019}, which also analyses the occurrence rates of the ultracool dwarfs observed by K2.}

The \textit{K2} mission is ideal for performing occurrence rate studies due to the simultaneous observation of thousands of stars, with long observation timescales. Unlike the original \textit{Kepler} mission, \textit{K2} has observed thousands of mid-to-late M dwarf targets \citep{howell2014,dressing2017_planets}. In this paper, we perform a full transit search of the \textit{K2}'s ultra-cool dwarf sample, using our own photometric and detrending pipeline to correct for \textit{K2}'s systematics and stellar variability. Using an injection-recovery scheme to calculate our detection sensitivity, we use our results to constrain the occurrence rate of planets orbiting ultracool stars.

% TODO: here I could add: We fit the occurrence rates within a hierarchical Bayesian model (HBM) to account for uncertainties in planet parameters. Additionally, we show a convenient way to simplify the HBM likelihood function, by directly marginalising over the uncertain planet parameters, and thereby greatly reducing the model parameters in cases with many detected planet candidates. Our method is applicable for occurrence rate studies in a general settings.

%As well as informing future observation strategies, the
% Using a different strategy, \citet{obermeier2015} observed a very large number of M dwarfs ($\sim 60000$ targets) with the Pan-Planets survey. The survey yielded an upper limit of 0.34\% on the occurrence rate of hot Jupiters around mostly mid- and early-M dwarfs.

% PARAGRAPH: What we will aim to study: planets in the habitable zone, earth-like, but also super-Earths. What are previous trends? Increasing planet occurrence rate with low stellar masses. However, gas giants also decreasing. Mention that gas giants have been found around M-Dwarfs (KOI 254.01, 096 Rj, Johnson et al 2012, reference at the bottom of Dressing 2013 page 2), can they be found around ultracools? Mention previous occurrence rates. Then to super-Earths: trend of increasing occurrence rate with decreasing stellar mass. Does this continue into the ultracools? What are previous results to compare to? At least mention papers.

% Greater detail in introducing occurrence rate study procedures; i.e what is completeness, what is the statistical problem, etc... To mention results?

We characterise the sample of targets and present our methods for creating lightcurves for each target in Sect. \ref{sec:data}. We describe our detrending and transit-search pipeline in Sect. \ref{sec:pipeline}, and the procedure for determining our pipeline's detection sensitivity in Sect. \ref{sec:detection-sensitivity}. In Sect. \ref{sec:model}, we explain our statistical model for inferring the occurrence rates with uncertain planet parameters, and we also show a convenient simplification of the resulting likelihood function. We present the inferred occurrence rates in Sect. \ref{sec:results} and discuss the implications in \ref{sec:discuss}.

%%%%%%%%%%%%%%%%%%%%%%%%%%%%%%%%%%%%%%%%%%%%%%%%%%%%%%%%%%%%%%%%%%%%%%%%%%%%%%%%

\section{Data} \label{sec:data}

% TODO: for style, switch to inferno colour scheme

\subsection{Target characterisation}

%
% The observation periods of our data range from 48 to 88 days

Our targets are taken from a number of Guest Observer programs focusing on ultracool stars and brown dwarfs, containing \res{1310} unique EPIC numbers. In several cases however, two EPICs were found to refer to the same target. After resolving all duplicates and performing the target selection described below, we are left with \res{702} unique characterised ultracool dwarfs for the occurrence rate estimation. Some of the stars have observations spanning multiple campaigns; for such cases, we take the detection sensitivity of the lightcurve with the best detrended noise characteristics (see Sect. \ref{subsec:detrending}). We did not include targets from Campaign 1 due to the weaker photometric quality, and Campaign 9 as it focused on microlensing.

We first characterise the target stars to find their radii and masses. We use these parameters to filter out our ultracool dwarf sample, and remove any mis-classified stars. Distant red giants can masquerade as bright ultracool stars, with the same surface temperatures and colour indices \citep{mann2012,dressing2017_stars}. Including these in our sample would significantly affect our results: red giants tend to be brighter and have more precise photometry than ultracool stars, which would cause us to overestimate our sample detection sensitivity. We can distinguish between red giants and cool dwarfs with the parallaxes provided by \textit{Gaia}.

We perform our own stellar characterisation, as the KIC and EPIC values have been found to underestimate radii for low mass stars by up to 39\% \citep{brown2011,boyajian2012,huber2016,dressing2017_stars}. Cross-referencing with \textit{Gaia} DR2\footnote{We use the cross-matched database created by Megan Bedell: \url{gaia-kepler.fun/}} \citep{marrese2019} and \textit{2MASS} \citep{skrutskie2006}, we find full 5-parameter astrometric solutions for \res{1143} individual EPIC identifiers, which we use to estimate distances to the stars. We then perform spectral energy distribution (SED) fitting on each target, with the combined \textit{Gaia} and \textit{2MASS} photometry, using the VO SED Analyser\footnote{\url{svo2.cab.inta-csic.es/theory/vosa/}} (VOSA) \citep{bayo2008,rodrigo2017}. We use the VO catalogues to also add photometric magnitudes from other sources, e.g. including \textit{WISE} \citep{wright2010}, \textit{SDSS} \citep{aguado2019}, \textit{UKIDSS} \citep{lawrence2007}, \textit{PAN-STARRS} \citep{chambers2016}, where available. We fit the SEDs with the BT-Settl, BT-COND, BT-DUSTY \citep{allard2012}, and BT-Settl-CIFIST \citep{baraffe2015}, selecting the highest-likelihood model with the VOSA tool. For each star, we thus obtain the stellar effective temperature, $T_{\text{eff}}$; and where the parallax is known, the luminosity, $\frac{L}{\si{\Lsun}}$.

We also cross-reference our target sample with the online census compiled by J. Gagné\footnote{\url{jgagneastro.wordpress.com/list-of-ultracool-dwarfs/},\\\url{jgagneastro.com/list-of-m6-m9-dwarfs/}}. This list contains well-characterised ultracool dwarfs and brown dwarfs taken from various sources in the literature, and also includes spectral types. Our SED-derived spectral types agree with the values in the Gagné census to within 2 spectral sub-types for 97\% of cases, and within 1 spectral sub-type for 88\% of cases. We also find that the difference in the stellar indices is symmetrically distributed, meaning that we are not biased to over- or under-estimating the stellar temperatures compared to previously-published values.

% As a first check of misclassified targets, we cross-reference our target sample with the online census compiled by J. Gagné\footnote{\url{jgagneastro.wordpress.com/list-of-ultracool-dwarfs/}}\footnote{\url{jgagneastro.com/list-of-m6-m9-dwarfs/}}. This list contains well-characterised ultracool dwarfs and brown dwarfs, taken from various sources in the literature, and includes estimated spectral types.

% To ensure we do not include any mis-classified red giants, we first cross-reference our target sample with the online census compiled by J. Gagné\footnote{https://jgagneastro.wordpress.com/list-of-ultracool-dwarfs/}\footnote{https://jgagneastro.com/list-of-m6-m9-dwarfs/}. This list contains well-characterised ultracool dwarfs and brown dwarfs, taken from various sources in the literature, and includes estimated spectral types. This leaves us with \res{467 targets}.

\begin{figure*}%
    \centering
    \includegraphics[scale=0.6]{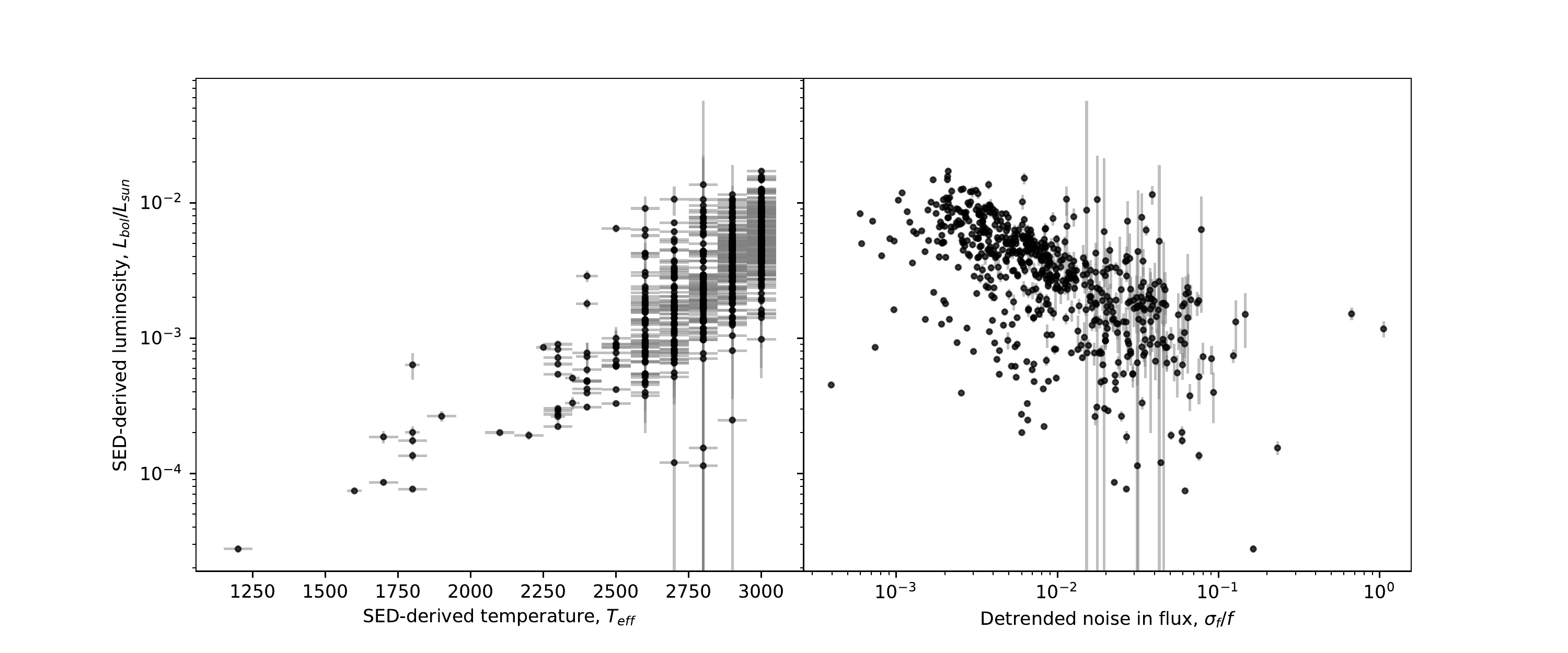}
    \caption{The distribution of stellar parameters for our filtered targets. Left: the SED-derived temperatures and luminosities. Right: the SED-derived luminosity and the fractional noise in the fully-detrended lightcurve of each target. Faint targets (with high noise in the lightcurve) have large uncertainties in their parallaxes and therefore their luminosities.}
  \label{fig:stellar_distribution}
\end{figure*}

% update which models I use

In selecting our targets, we first cut out all targets with SED-derived $T_{\text{eff}} > 3000K$ and $R_{star} > 0.5 \si{\Rsun}$. The high radius cut allows us to keep faint stars with poor precision on their parallax measurements (see Fig. \ref{fig:stellar_distribution}). We find a number of targets with very imprecise parallaxes; \res{213} EPIC numbers have a parallax-over-error of less than 10. For these targets, the stellar radii are very uncertain and biased to higher values.

A number of faint targets present in the J. Gagné list are also missing parallaxes, or were not found in \textit{Gaia} DR2. We cannot directly infer $\frac{L}{\si{\Lsun}}$ for these stars, but the majority of targets with missing parallaxes have very faint magnitudes - clustered around a \textit{Gaia} G-band magnitude of 20. For the \res{20} targets with \textit{Gaia} data but missing astrometry, the mean and brightest G magnitudes are 20.1 and 16.7 respectively, while the mean and brightest \textit{2MASS} J-band magnitudes are 15.9 and 12.4 respectively. For another \res{94} targets that couldn't be cross-matched to \textit{Gaia} DR2, we find mean and brightest \textit{2MASS} J magnitudes of 16.0 and 11.1 respectively. All of these targets are characterised as ultracool or brown dwarfs in the J. Gagné census, and are unlikely to be mis-characterised red giants due to their low magnitude \citep[see][]{mann2012}. We therefore include them in our target list.

%These are unlikely to be red giants \citep{mann2012}, we do not include them in our sample, even if they are in the Gagné census.}

We also check the detrended noise of each target as a function of its \textit{Gaia} G magnitude (see Sect. \ref{subsec:detrending} for the detrending procedure). A number of targets have a noise-level below what we would expect based on the magnitude, which could be due to the target's PSF being blended with a brighter star. We filter out \res{87 such targets.}

It should be noted that TRAPPIST-1 is not a part of our stellar sample. While it was included in a Guest Observer proposal for campaign 12 - before the discovery of its planets - the star itself would have fallen just outside one of the CCD frames in the initial observation field. Only after the discovery of the planets by \citet{gillon2016} was the campaign 12 field modified to include TRAPPIST-1. The presence of TRAPPIST-1 in our sample would therefore be conditioned on it having an already discovered system. Without modelling for this selection bias, we cannot include TRAPPIST-1 in the sample.

% NO LONGER RELEVANT: To this, we also add \res{77} targets from the characterised census compiled by J. Gagné, with SED-derived $T_{\text{eff}} < 3000K$ but very imprecise parallax measurements. These \res{77} targets all have Gaia magnitudes of $G > 16$, mostly clustered around $G \sim 20$, which makes them unlikely to be red giants \citet{mann2012} \tocite \todo{Need more concrete citations on this}.} \qbrice{My argument here is this: Mann 2012 finds that above Kpmag 14, only 7\% of cool targets are misclassified as giants. Now here, we use the cutoff of 16, but instead of the \textit{Kepler} mag we use the Gaia mag. Is this still solid enough?}

To calculate transit depths, we must input values for stellar radii and masses. Targets with imprecise parallaxes have large uncertainties in stellar radii; in some cases, the means of the radii are inconsistent with ultracool dwarf isochrones and evolutionary tracks. We therefore use the following treatment: we use our SED-derived $T_{\text{eff}}$ to match to a spectral type, and take the corresponding stellar radius from the table compiled by E. Mamajek\footnote{\url{pas.rochester.edu/~emamajek/EEM_dwarf_UBVIJHK_colors_Teff.txt}} \citep{pecaut2013}. For targets with no parallaxes, we use the spectral type quoted in J. Gagné instead. In total, our filtered catalogue contains \res{702} unique targets, with \res{644} M Dwarfs, \res{46} L Dwarfs and \res{3} T Dwarfs.

\subsection{Lightcurve photometry}

For each target, we download the long-cadence ($\sim30$ minute exposure) target pixel files (TPFs), and perform the photometry ourselves, rather than using the lightcurves processed by the \textit{Kepler/K2} SOC. For dim targets, we find that using our custom pipeline improves our final photometric precision over the \texttt{PDC} lightcurves, and this also allows us to better control our systematic noise.\footnote{Comparing our lightcurves with the \texttt{PDC} lightcurves, we find that some of the \texttt{PDC}s contain anomalously high noise and potentially altered flux variations. This was noticed for TRAPPIST-1, where both the noise and the transit depths in the \texttt{PDC} lightcurve are inflated by a factor of $\sim 5$. It's possible that the K2SOC pipeline isn't well-suited for very faint targets.}

Some of the targets are referenced by multiple EPIC numbers, with separate TPFs for each within the same campaign (usually on the order of 10 arcseconds apart). When this occurs, we use the \textit{Gaia} proper motions to estimate the position of the target during the observation, and take the EPIC with the closest listed coordinates.

Due to its failed reaction wheels, \textit{K2}'s pointing angle is constantly drifting and being corrected by thruster firings \citep{howell2014}, meaning that the targets move across their TPFs over $\sim6$ hour timescales. To perform the photometry, we use circular top-hat apertures following each target, with a 2 pixel radius. To begin with, we search within 1.5 pixels of the predicted position to lock on to the target, using the WCS header in each TPF to transform astrometric into pixel coordinates. We repeat this procedure on 100 randomly selected frames to find the median offset from the expected position. Each TPF provides estimated pointing offset coordinates for each frame, in \texttt{XPOS\_CORR} and \texttt{YPOS\_CORR}, in units of pixels. To ensure the aperture follows the target, we offset it by \texttt{XPOS\_CORR} and \texttt{YPOS\_CORR} in each frame.

Finally, we remove points that are next to thruster-firing events, as well as other points flagged in the TPF \texttt{QUALITY} column \citep{jenkins2010}.  We find that our photometric procedure already removes a significant portion of the systematic noise present in the \texttt{PDC} lightcurves.

%%%%%%%%%%%%%%%%%%%%%%%%%%%%%%%%%%%%%%%%%%%%%%%%%%%%%%%%%%%%%%%%%%%%%%%%%%%%%%%%

\section{Transit-search pipeline} \label{sec:pipeline}

% Subsections: detrending, transit search and fitting, manual vetting
% Introduce the term detrending, bls etc... explicitly

To estimate occurrence rates, we need to calculate the completeness of our search; meaning the probability of finding a planet for a particular target, if it were there. We must first create an automatic pipeline which encompasses all the data processing, including the removal of noise and the subsequent transit search. As our target stars are faint, which limits our photometric precision, we must make every effort to maximise the quality of our data and remove as much noise as possible. The entire pipeline must be computationally efficient, as we will repeat it thousands of times per target (see Sect. \ref{sec:detection-sensitivity}).

\subsection{Detrending} \label{subsec:detrending}

\begin{figure}
    \centering
    \includegraphics[width=\hsize]{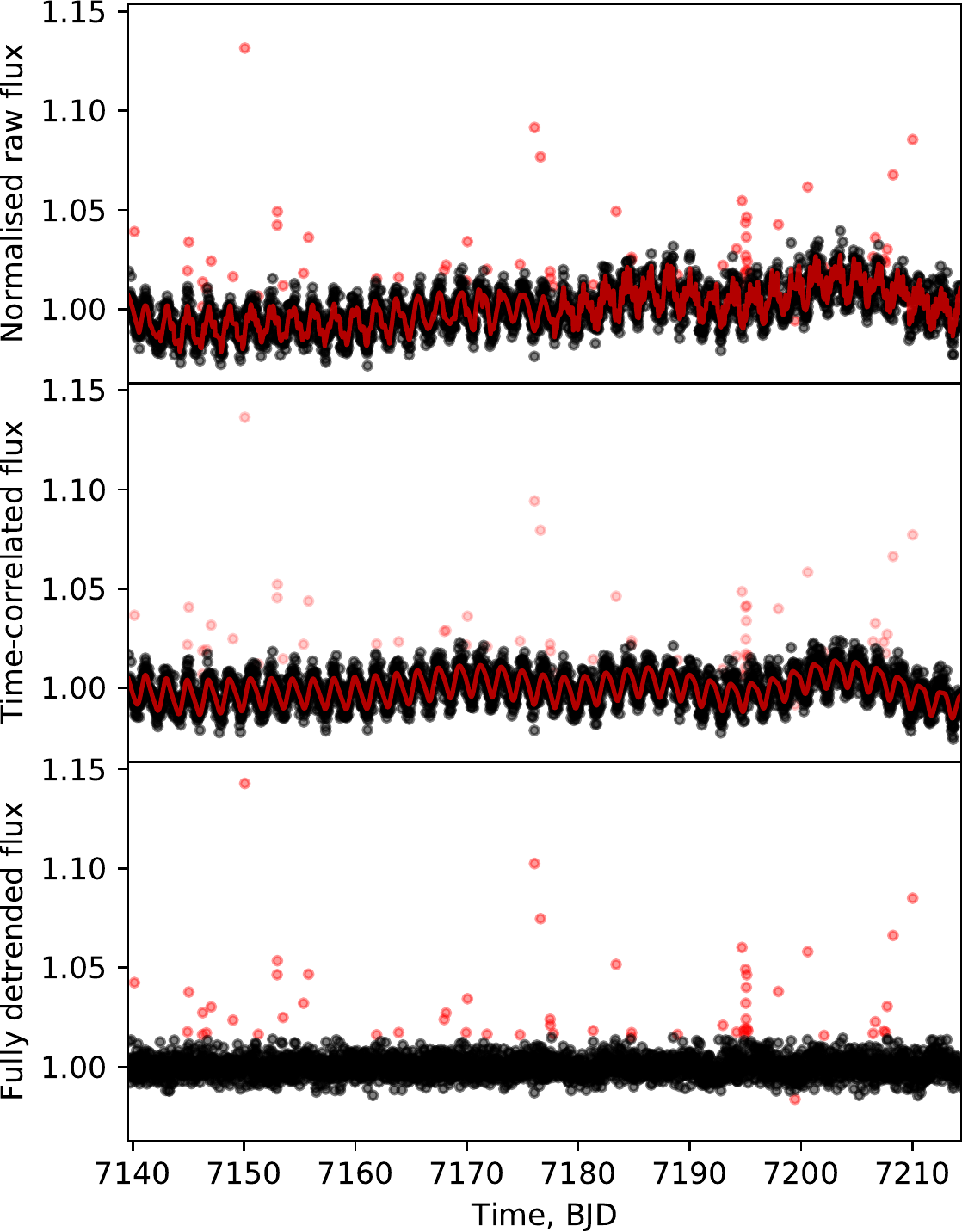}
        \caption{An example of the detrending procedure for target EPIC 212035051 in campaign 5. The improvement in the flux standard deviation is from $0.008$ for the raw lightcurve, to $0.004$ for the detrended lightcurve. \textit{Top:} raw flux; \textit{middle:} time-correlated component, mostly quasiperiodic stellar variability, with pointing drift noise removed; \textit{bottom:} detrended lightcurve on which we perform the transit search. \response{Red points are outliers at more than 3 standard deviations from the trend. They are masked out during the detrending; i.e. we do not use them to predict the noise trend.}}
    \label{fig:detrend-plot}
\end{figure}

The lightcurves produced by \textit{K2} photometry are well-known to suffer from correlated systematic noise. This is caused by the unstable pointing vector of the spacecraft, which drifts on 6-hour timescales and needs to be re-positioned with thruster firings to keep it on target. As each pixel has an (unknown) pixel-response-function (PRF), this drift introduces noise into our photometric lightcurve, which can be larger than the transit signals we are looking for. As a consequence, the pointing-drift noise is strongly correlated with the $x$ and $y$ positions of the star on the TPF. Additionally, many stars exhibit stellar variability, producing either a long-term and smooth variation in flux, or shorter timescale variations which are generally quasi-periodic.

Most transit-search algorithms require a flattened lightcurve, and the systematic noise can mimic transit signals at short timescales. Removing both sources of systematic noise while preserving potential transit dips is a crucial step in our data analysis pipeline.

To that end, we use Gaussian process (GP) regression to fit for both sources of noise simultaneously, allowing us to remove them (referred to as detrending). Gaussian processes are a popular way to remove systematic signals from \textit{K2} data \citep[see][on which our procedure is based]{aigrain2016}. We only present a brief outline here.

Our GP model uses three predictors: the time, $t$, as well as the $x$ and $y$ positions of the star on the TPF. The basic version of our model uses decaying squared-exponential kernels in each of those parameters:

\begin{equation}
    K_{ij} = k_{xy}(x_i, y_i, x_j, y_j) + k_t(t_i, t_j) + \sigma ^2 \delta_{ij}
    \label{eq:gp_kernel_1}
\end{equation}

where $i,j$ index two different points in the lightcurve and $\delta_{ij}$ is the Kronecker delta function, and where:

\begin{eqnarray}
    k_{xy}(x_i, y_i, x_j, y_j) & = & A_{xy} \exp \left( -\frac{(x_j - x_i)^2}{L_x^2} -\frac{(y_j - y_i)^2}{L_y^2} \right) \\
    k_t(t_i, t_j) & = & A_t \exp \left( - \frac{(t_i - t_j)^2}{L_t ^2} \right)
\end{eqnarray}

\response{On a subset of lightcurves, we also tested the performance of the Matérn 5/2 kernel, which places fewer restrictions on the fit's smoothness. The average difference in performance wasn't significant, though in the future it may be worth selecting the kernel on a case-by-case basis, depending on which shows better detrending performance for a given lightcurve.}

Before conducting the full detrending procedure, we perform a short run and calculate a Lomb-Scargle periodogram of the time-correlated component of our lightcurve. If a periodicity is detected above a certain threshold, we instead use a quasi-periodic kernel for the time parameter:

\begin{equation}
    k_t(t_i, t_j) = A_t \exp \left( - \frac{(t_i - t_j)^2}{L_t ^2} - \Gamma \sin ^2 \frac{\pi (t_i - t_j)}{P} \right)
\end{equation}

Our pipeline is re-purposed from the one used in \citet{luger2017,grimm2018}, with modifications for the increased speed required here. We perform four consecutive hyperparameter fits in a process similar to sigma-clipping, to flag outliers that will be ignored in subsequent GP fits. Masking the outliers is only relevant during the detrending stage; they are not removed for the transit search. During the hyperparameter-fitting procedure, we use a subset of the flux measurements to calculate the likelihood at each iteration. We only use the full lightcurve for the predicting the noise components, i.e. during the outlier flagging and during the final detrending. We show an example of the detrending results in Fig. \ref{fig:detrend-plot}.

Having fit the systematic noise, we subtract it out to produce flattened lightcurves, on which we then perform a transit search. For many targets, we also notice a ramp-like feature in the beginning of the lightcurve. We fit this simultaneously with the GP hyperparameters using an exponential function, and subtract it out.

% TODO: analysis of the detrending results; quasi-periods etc... Put some numbers up actually.

The Gaussian process regression is implemented using the \texttt{george} package in \texttt{python} \citep{hodlr}.

\subsection{Transit search} \label{subsec:transit-search}

% TODO:
% I need to actually check the algorithm again and write it back down in detail.
% Limb-darkening parameters
%

We then perform a standard transit-search; see e.g. \citet{dressing2015,vanderburg2016,mayo2018} for similar procedures. We first calculate a series of box-fitting-least-squares (BLS) spectra, to find periodic dips in our data \citep[][]{kovacs2002}. We modify the BLS algorithm to only search for dip durations shorter than the maximum transit duration at a particular period, with a 3x tolerance factor. We also limit the search resolution to the minimum required for a particular period, with 2x tolerance. These steps greatly speed up the BLS procedure.

At each iteration, we note the peak in the BLS spectrum, corresponding to the "strongest" periodic dips \citep[see][]{kovacs2002}. We then remove the points within two durations of each dip, before rerunning the BLS algorithm to search for more signals. We test each peak by removing the 2 points with the lowest flux; if this changes the average depth of the peak by more than 50\% we flag the peak as invalid. Peaks with negative depths are also flagged as invalid. This is repeated until 10 valid peaks are found.

We then perform a physical transit model fit on each valid BLS peak. The model used is based on \citet{mandel2002} and implemented in the \texttt{batman} package \citep{kreidberg2015}. From this we obtain joint distributions for the planet radius, period, first-transit time $t_0$ and impact parameter. We assume zero eccentricity, as the planets we search for are all in close-in orbits.

For each fit, we calculate a signal-to-noise ratio (SNR), defined as

\begin{equation}
    SNR = \frac{d\sqrt{N_{p}}}{\sigma}
\end{equation}

where $d$ is the transit depth, $N_p$ is the number of points in transit, and $\sigma$ is \response{estimated in the lightcurve's fitted GP model (see Eq. \ref{eq:gp_kernel_1}). $\sigma$ is therefore the sigma-clipped standard deviation of the detrended lightcurve.} Valid BLS peaks that produce an SNR greater than $\snrlim$ are passed to visual inspection. From visual inspection, we can discard peaks that are clearly noise or anomalies, and keep peaks that could potentially be transits. Obvious and common systematic noise includes points near thruster firings, displaying a clear ramp going into the thruster firing event. This also produces a very asymmetric dip. Occasionally, we also noticed temporary bouts of heavily increased white noise, where a few points below the flux median could produce a high-SNR dip.

\subsection{Transit-search results}

We test our detrending and transit-search pipeline on TRAPPIST-1, finding 6 of the 7 planets. The last planet, TRAPPIST-1h, only has 4 transits in the \textit{K2} data, with one occurring during a flare, and one overlapping with a transit from planet $b$ \citep{luger2017}. As the smallest planet in the system, its discovery was very near \textit{K2}'s detection limit. Without using prior knowledge to search for it - as was done when possible periods and ephemerides were predicted from Laplace resonances in \citet{luger2017} - it's unsurprising that it cannot be found.

Performing the search across our entire sample of targets, our pipeline identifies \res{116} potential transit signals. We examine these visually, and we also perform a search for possible aliases by focusing on alternate peaks in the BLS spectrum. The majority of signals near our SNR limit are caused by residual noise in the lightcurve. This is mostly caused by stellar variability below our detection limit, limited sections of the lightcurve with increased noise, or instances where the star drifts off the TPF aperture. Some lightcurves also have leftover red-noise at timescales below what we detrended.

% Potentially: EPIC 211101996 shows up with a transit but it's a blended PSF (transit duration is too long, lightcurve noise way too low).

Out of the filtered signals, we find 5 credible transit-like signals. Three of them are eclipsing binaries which have been previously identified in the literature. Their EPIC numbers are 211079188 \citep{kruse2019}, 212002525 \citep{gillen2017}, and 211075914 \citep{david2016}. We find one clear planetary signal in EPIC 210490365, which corresponds to  K2-25b. Our retrieval produces a radius of $1.7\pm0.2\si{\Re}$. \citet{mann2016} identified this as a Neptune-sized planet, with radius $3.42^{+0.95}_{0.31}\si{\Re}$ and period $3.43$ days. \citet{mann2016} use the star's membership in the Hyades' cluster to estimate its distance, which they use to calculate the luminosity. For the radius/mass estimate however, they use the relations derived in \citet{mann2015}; the estimates are significantly larger than the model values we use based on an M5.5V dwarf (see Sect. \ref{sec:data}). For the occurrence rate calculation, we adopt their published values and uncertainties for the radius and period, which include their stellar radius uncertainty.

%%%%%%%%%%%%%%%%%%%%%%%%%%%%%%%%%%%%%%%%%%%%%%%%%%%%%%%%%%%%%%%%%%%%%%%%%%%%%%%%

\section{Detection sensitivities} \label{sec:detection-sensitivity}

% On eccentricity, perhaps quote Youdin or something about the effect of eccentricity on detectability (optional).
%
% The uncertainty in stellar mass and radius comes here: I need to explain how I treat it here, so that I don't have to do it later.
% Introduce completeness symbol (find out what the conventional symbol is). Would that lead to too many symbols?
% Introduce the symbols for detection sensitivity and geometric transit probability.
%
% In fact: introduce the geometric transit probability here; as I will need to show how I calculate the completeness within the bins (using the geometric transit probability
% a "weighting" for each injected point.

\begin{figure*}%
    \centering
    \subfigure[Detection sensitivity]{ %
        \label{fig:detection-sensitivity}
        \includegraphics[scale=0.48]{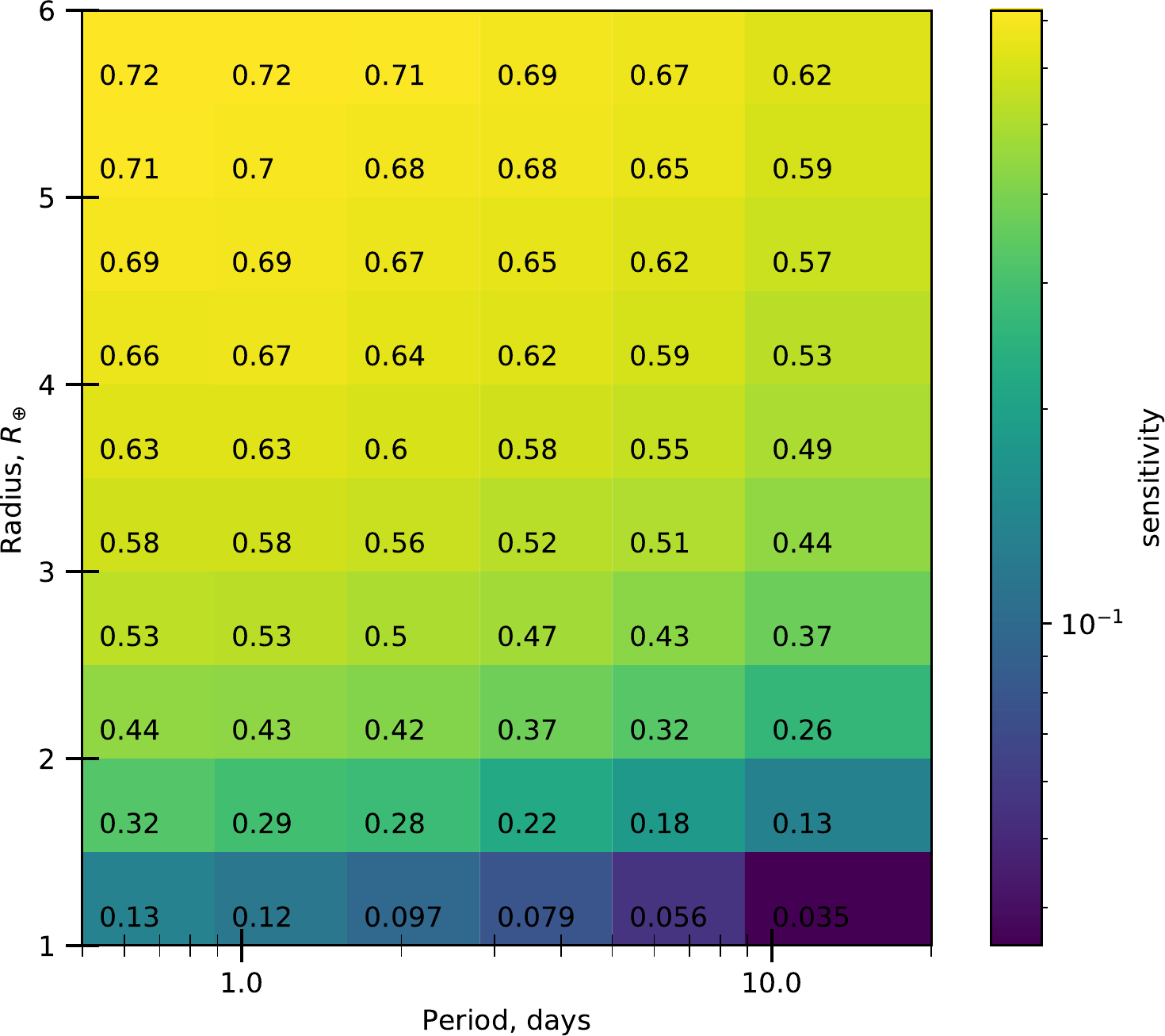}
    }
    \qquad
    \subfigure[Completeness]{
        \label{fig:completeness}
        \includegraphics[scale=0.48]{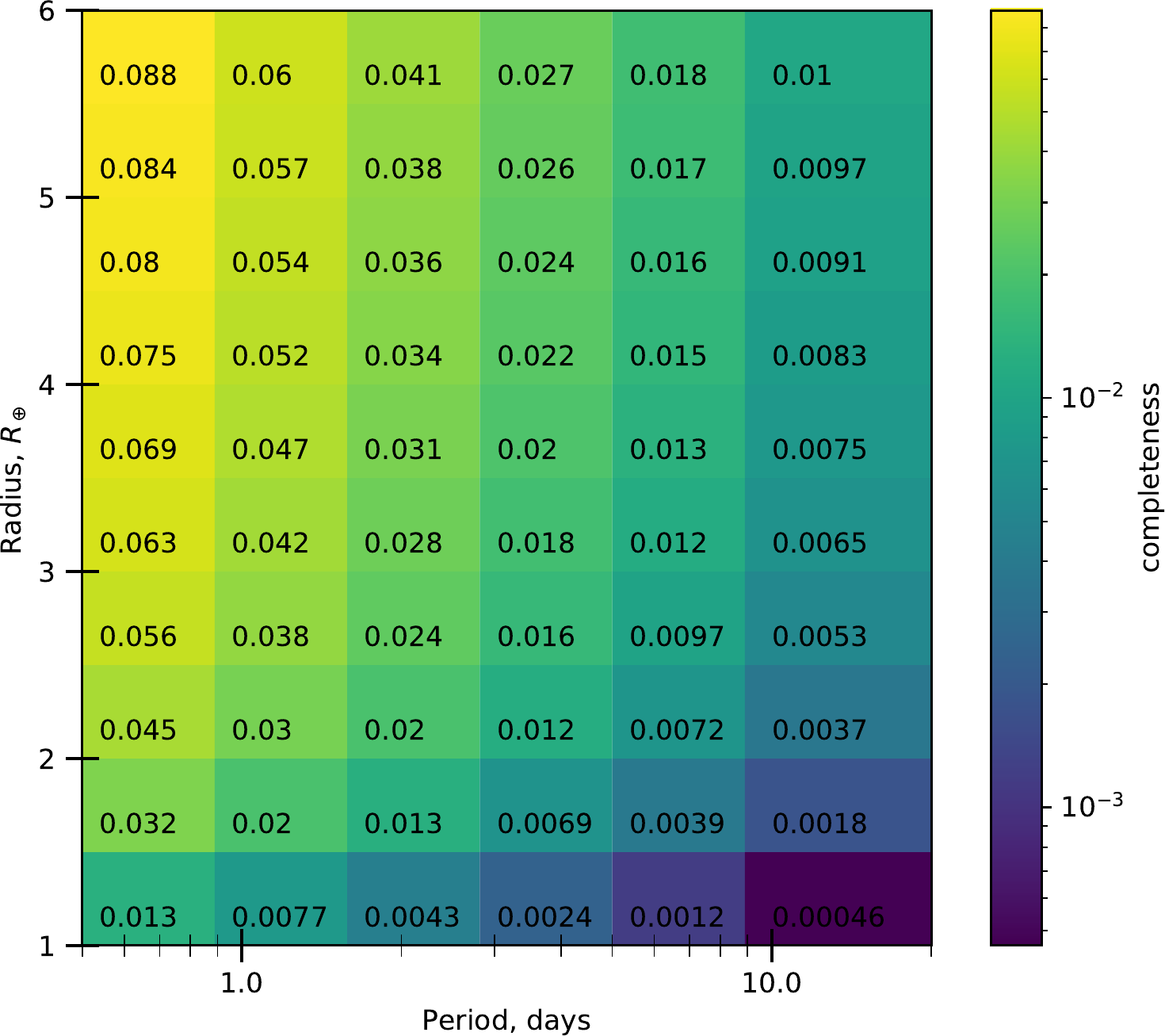}
    }
  
%   \begin{subfigure}{0.49\textwidth}
%     \centering
%     \includegraphics[scale=0.48]{figures/sensitivity_grid.pdf}
%     \caption{\ref{fig:HP}a: Detection sensitivity}
%     \label{fig:detection-sensitivity}
%   \end{subfigure}
%   \hfill
%   \begin{subfigure}{0.49\textwidth}
%     \centering
%     \includegraphics[scale=0.48]{figures/completeness_grid.pdf}
%     \caption{\ref{fig:HP}b: Completeness}
%     \label{fig:completeness}
%   \end{subfigure}
  
  \caption{Left: the detection sensitivity, showing the chance of finding a planet in the data, given that it's transiting. Shown as a function of planet radius and orbital period. Right: the completeness, showing the chance of finding a planet with a random orbital alignment. In both cases, we integrate over the entire sample of \res{702} stars, thus showing the mean of the sensitivity/completeness across our entire sample. For example, if TRAPPIST-1b (R=1.086, P=1.51d, \citet{gillon2017}) were orbiting a random star in our sample, we would have a \res{0.77\%} chance of finding it given a random inclination, or an \res{12\%} chance of detecting it if it was transiting.
  }
  \label{fig:irm-results}
\end{figure*}

In order to estimate the occurrence rates of planets from our data, we need to know our detection sensitivity as a function of the planet parameters (i.e the chance of finding a planet given that it transits). Each lightcurve has a different noise profile, and the most accurate way to find our detection sensitivity is to directly test our transit-search pipeline on the real lightcurves themselves. We perform injection-recovery modelling, injecting 1000 synthetic transit signals per lightcurve, with random planetary/orbital parameters, using \texttt{batman} \citep{kreidberg2015}. Each injected synthetic signal has planetary radius, orbital period, inclination and eccentricity ($R_p$, $P$, $i$, $e$) taken from the following distributions:

\begin{eqnarray*}
\frac{R_p}{\si{\Re}} & \sim & U(0.5, 6.00)\\
\log \frac{P}{\si{day}} & \sim & U(\log(0.5), \log(20))\\
t_0 ~ & \sim & U(0, P)\\
i & \sim & U(i_{min}, i_{max})\\
e & = & 0.0\\
\end{eqnarray*}

where $i_{max}$ and $i_{min}$ are calculated for each star as the maximum and minimum inclination for a transit, and where $~ U(a, b)$ signifies that a "drawn from a uniform distribution between $a$ and $b$." We fix the eccentricity at $0.0$, as we do not know the underlying eccentricity distribution. This assumption has a small effect; close-in planets around ultracool dwarfs are subject to strong orbital circularisation \citep[see e.g.][]{luger2017}. \response{The distribution of injected radii and periods is taken to match the distribution of bins we use to parametrise the detection efficiency and occurrence rate (see Sects. \ref{sec:model} and \ref{sec:occr-unconstrained}).}

We then run our entire transit-search pipeline on each modified lightcurve, and attempt to retrieve the transit parameters. This includes the detrending stage, as it can in principle remove or smear out transit signals. In order to count as a detection, we have the following tolerances: the detected period must match to within 1\%, the time of first transit must match to within half a transit duration, and the duration must match to within 50\% of the injected duration. Period aliases of the injected signal are treated as detections (though this naturally reduces their calculated SNR, which may lead to them being rejected). \response{We look for alias periods at factors of $\frac{1}{3}, \frac{1}{2}, 2, 3$ of the injected period, with the same tolerance criteria as normal signals.} We find an average of around 3 alias detections per 1000 injections.

\response{We split the $R_p$ and $\log P$ space into a set of $N_B$ bins, $\{B_j\}_{j=1}^{N_B}$, and count the fraction of injections that were recovered in each bin. There is a trade-off between the size and number of bins we use, and the number of injections we need to make per lightcurve. With smaller bins, the distribution of injected radii and periods within each bin is less important, and we better approximate the true smooth nature of the planetary occurrence rate distribution. To maintain the same level of per-bin sampling, however, decreasing the bin size requires an increase in the number of injected signals per lightcurve by the same factor.}

% Last count on aliases: ~ 500 per 166000 injections, so about a 3 in 1000 rate.

We do not include visual identification as a step in the injection-recovery modelling, due to the sheer number of lightcurves produced. This is a potential source of bias, and is equivalent to assuming that we wouldn't incorrectly identify a real transit as noise, if it got past all the stages of our transit search. This is also why we had to make our pipeline strongly automated, capable of discarding as much noise as possible on its own: to reduce our reliance on visual identification.

As a result, we must be careful in Sect. \ref{sec:pipeline}, in deciding which features to consider noise. The rejected signals must have a feature that clearly sets them apart from a physical transit. If we remove transit-like signals that could be produced by our physical model, then we may also remove actual transit signals. Moreover, we would have to apply the same selection criterion in removing injected signals. And as we do not visually inspect the injected signals, this would bias our results.

The fraction of injected planet signals that are detected in a particular lightcurve, as a function of planet period and radius, is taken as the detection sensitivity in that bin, \ds. A plot of our sample-wide detection sensitivity is shown in Fig. \ref{fig:detection-sensitivity}, integrated over the entire sample of targets.

For a given planet injection, we can also calculate the geometric transit probability, given by

\begin{equation}
    \tr = \frac{R_{\star} + R_p}{a}
    \label{eq:geometric_transit}
\end{equation}

where $a$ is the orbital semi-major axis, and $R_\mathrm{\star}$ is the radius of the star.

Finally, for a particular star and within a particular bin, $B_j$, we approximate the average completeness, $\comp_j$. We take the completeness as the number of detected injections weighted by by the transit probability of each injection, and divide by the total number of injections falling inside the bin, $N_j$:

\begin{equation}
    \comp _j = \frac{1}{N_j} \sum _{l=1}^{N_j} \mathds{1}_l \cdot \tr{}_{,l}
    \label{eq:completeness}
\end{equation}

\response{where $\tr{}_{,l}$ is the geometric transit probability of the $l^{\text{th}}$ injection, and} where

\begin{equation}
    \mathds{1}_l=
    \begin{cases}
        1 & \text{$l^{\text{th}}$ injection detected,} \\
        0 & \text{otherwise.}
    \end{cases}
\end{equation}

% \todo{Find external examples of where this is used to refer to as backup.}

%%%%%%%%%%%%%%%%%%%%%%%%%%%%%%%%%%%%%%%%%%%%%%%%%%%%%%%%%%%%%%%%%%%%%%%%%%%%%%%%

\section{Modelling the occurrence rate} \label{sec:model}

% Regarding symbol convention, there are two options:
%   1. Capital letters for integrated/mean values for the whole sample (i.e N_exp, L, H), minuscule letters for values related to a single star (i.e eta, n_exp, l)
%   2. Overbar to refer to values that are integrated/mean for the whole sample (i.e over N_exp, L, H). This makes more sense I believe, but then I'll need to make some other changes. Firstly, I will need to figure out another way to differentiate between the Poisson likelihood and the total model likelihood (perhaps L_poisson). There may be other problems, need to figure it out.
%
% TODO: consider using the machine learning convention that indexes over samples in the data/training-set are indexed by a super-scripted index in parentheses.
%
% TODO: in the HBM section, potentially reorder: 1. introduce HBM and both the joint posterior, and the marginalised likelihood in occurrence rates (complication: if I do this, I will need to start from the posterior/Bayes inference, since I'll need to show where the non-marginalised likelihood comes from); 2. explain that we show here a way to solve this integral directly, without needing to sample/infer the posteriors for the true planet parameters; 3. start with converting the poisson likelihood into the form without bins; 4. do the rest of the derivation.
%
% Show that prior shouldn't affect the upper-limits (especially if I place a lower-bound cutoff).
% Analytic derivation (or at least a simplified single-step numerical derivation) of the upper-limits on the occurrence rate.

We treat planet detections as being drawn from an inhomogeneous Poisson point process. In other words, for the $i^{\text{th}}$ star, the detection of a planet with radius and orbital period $(R_p, P)$ is an event that occurs with a rate $\lambda _i$ which depends on $(R_p, \log P)$\footnote{We explicitly parametrise by $\log P$ as our completeness function bins have uniform intervals in $\log P$, and the completeness function varies more gradually when parametrised by $\log P$. The completeness function varies much more strongly between $1.0-1.5$ days than it does between $20.0-20.5$ days for example.}. We will henceforth denote our planet parameters with $\boldsymbol{x} = (R_p, \log P)$.

The rate density of planet detections, $\lambda_i (\boldsymbol{x})$, is the result of several probabilistic steps. Firstly, there is a true planetary occurrence rate, \occr, which determines the number of planets per star in our sample, including non-detections. This is the unknown quantity we seek to find, and we take it to be the same for all the stars in our sample. Of the planets that may orbit the stars in our sample, only a fraction will have orbits aligned so that they transit from our point of view. We refer to this fraction as the geometric transit probability, $\tr$ (see Eq. \ref{eq:geometric_transit}). Finally, only a fraction of the transiting planets will be detected by our pipeline. Depending on the size of the signal and the number of transits, there is a chance our pipeline cannot distinguish it from noise. Included in this is the chance that transits fall within gaps in the data. We calculate this detection sensitivity, $\ds$, through injection-recovery modelling for each star (see Sect. \ref{sec:detection-sensitivity}). We combine the detection sensitivity and transit probability into the completeness, $\comp = \tr{} \ds{}$.

The total rate of planet detections is then given as the product of these probabilities $\lambda _i = \occr \comp_{i}$ for each star indexed by $i$, where each of those quantities depends on $(R_p, \log P)$. Once we can calculate $\tr{}_{,i}$ and $\ds{}_{,i}$, and thus $\comp_{i}$, we can use our data (detections and non-detections) to infer the occurrence rate, \occr.

% This needs a bit of refinement
To make the problem computationally tractable, we split our parameter space, $\boldsymbol{x} = (R_p, \log P)$, into a set of $N_B$ bins: $\{B_j\}_{j=1} ^{N_B}$. Treating the occurrence rate as constant within a bin, the parameters we seek to infer are the occurrence rates within the individual bins: $\bm{\occr} = \{\occr\}_{j=1}^{N_B}$. We then also treat the completeness as constant within each bin, and equal to the average completeness we calculate in Eq. \ref{eq:completeness}.

% Formally, the bins are disjoint, bounded Borel-measurable sets; i.e they do not overlap and have finite cardinality. And in the following equations, they need to cover our entire parameter space.

In the following sections, we present the likelihood functions we will use to infer the occurrence rate parameters, $\bm{\occr}$. We first do it for the case where detected planets' parameters are precisely known (without uncertainties), and then extend that to working with uncertain planet parameters.

\subsection{Precisely-known planet parameters} \label{subsec:precise-model}

The likelihood of a Poisson point process is easily defined when we know the precise values of events drawn from it, where events correspond to planet detections. For the $i^{\text{th}}$ star, we count the number of real planet detections in bin $B_j$ and denote it $d_{ij}$. For an inhomogeneous Poisson process on this parameter space \citep[see e.g][]{daley2007}, with a rate function $\lambda (\boldsymbol{x})$, the likelihood function for set of detections around the $i^{\text{th}}$ star is:

\begin{equation}
L_i = \prod_{j=1}^{N_B} \frac{(\Lambda_i (B_j))^{d_{ij}}}{d_{ij}!}\exp \left(-\Lambda_i (B_j) \right)
\label{eq:poisson-general-likelihood}
\end{equation} \\

where the measure $\Lambda (B_j)$ is the integrated rate in bin $B_j$, defined as:

\begin{equation}
    \Lambda_i (B) = \int _{B} \lambda_i (\boldsymbol{x}) d\boldsymbol{x}
\end{equation}

Within our framework, as the occurrence rate and completeness are taken to be constant within each bin, we can simplify $\Lambda (B_j)$ to

\begin{equation}
    \Lambda_i (B_j) = \occr_j \comp_{ij} |B_j|
\end{equation}

% volume -> cardinality is more accurate, but keep it for now

where $|B_j|$ is the volume of bin $B_j$, and $\comp_{ij}$ is the completeness in bin $B_j$, for the $i^{\text{th}}$ star. 

Assuming that the data for each star is independent, our total likelihood, for all $N_{\star}$ stars in our sample, is then:

\begin{align}
    L(\bm{\occr}) & = \prod_{i=1}^{N_{\star}} L_i \\
      & = \prod_{i=1}^{N_{\star}} \prod_{j=1}^{N_B} \frac{(\Lambda (B_j))^{d_{ij}}}{d_{ij}!}\exp \left(-\Lambda_i (B_j) \right) \\
      & = \prod_{i=1}^{N_{\star}} \prod_{j=1}^{N_B} \frac{(\occr_j \comp_{ij} |B_j|)^{d_{ij}}}{d_{ij}!}\exp \left(-\occr_j \comp_{ij} |B_j| \right) \label{eq:likelihood_star_product}
\end{align}

We only need to know the likelihood up to a multiplying constant, so we can remove any factors that do not depend on $\occr$, and we can further simplify the likelihood to obtain

\begin{align}
    L(\bm{\occr}) & \propto \prod_{j=1}^{N_B} \occr_j^{\sum_{i=1}^{N_{\star}} d_{ij}} \exp \left( -  \occr_j |B_j| \sum_{i=1}^{N_{\star}} \comp_{ij} \right) \\
    & = \prod_{j=1}^{N_B} \occr_j^{D_j} \exp \left( -  \occr_j N_{\star} \bar{\comp} _j |B_j| \right)  \\
    & = \left[ \prod_{j=1}^{N_B} \occr_j^{D_j} \right] \exp \left( - \sum_{j=1}^{N_B} \occr_j N_{\star} \bar{\comp} _j |B_j| \right)
    \label{eq:likelihood_precise}
\end{align}

where $D_j = \sum_{i=1}^{N_{\star}} d_{ij}$ is the total number of real planet detections in bin $B_j$, across all the stars in our sample, and $\bar{\comp}_j$ is the mean completeness in bin $B_j$, across our whole sample. $\bar{\comp}_j$ is the quantity plotted in Fig. \ref{fig:completeness}, and can be written:

\begin{equation}
    \bar{\comp}_j = \frac{1}{N_{\star}} \sum_{i=1}^{N_{\star}} \comp_{ij}
\end{equation}

The expectation of the total number of planet detections is equal to the term in the exponential of Eq. \ref{eq:likelihood_precise} \citep[see][]{youdin2011}:

\begin{equation}
    N_{\text{exp}} = \sum_{j=1}^{N_B} \occr_j N_{\star} \bar{\comp} _j |B_j|
    \label{eq:N_exp}
\end{equation}

$N_{\text{exp}}$ will also appear in the likelihood we derive for uncertain planet parameters.

% So we can also write the likelihood for precisely-known planet parameters as:

% \begin{align}
% L(\bm{\occr}) & \propto \left[ \prod_{j=1}^{N_B} \occr_j^{D_j} \right] \exp \left( - N_{\text{exp}} \right)
% \end{align}

\subsection{Planet parameters with uncertainties} \label{subsec:uncertain-model}

% TODO: I think this entire section might need to be reorganised a bit. First introduce the HBM and the idea behind it. Then do the poisson likelihood for x. Then put it all together.

The likelihood in Eq. \ref{eq:likelihood_precise} assumes that the planet radii and periods are known precisely, without uncertainties. This is not the case for a real catalogue of detections. To correctly treat uncertain planet parameters in our Bayesian formulation, we must use a hierarchical/multi-level Bayesian model (HBM). In the first application of an HBM to modelling planetary occurrence rates, \cite{foreman-mackay2014} show that an HBM recovers the underlying parameters of simulated data with a higher accuracy than traditional methods. We point the reader to \cite{foreman-mackay2014} for an explanation and justification of the principles of using a hierarchical framework to model planetary occurrence rates, and its advantages over traditional methods \citep[see also e.g.][for a more general introduction of HBMs in astrophysics]{hogg2010}.

In this section, we formulate our HBM, and show a convenient way to write the likelihood function in terms of only the parameters we are interested in (the occurrence rates, $\boldsymbol{\occr}$). We can therefore sample the parameters of our HBM as if it were a two-level model. Starting with the general likelihood equation for a Poisson point process, for the $i^{\text{th}}$ star, we can rewrite Eq. \ref{eq:poisson-general-likelihood} as:

%In brief, the idea behind using an HBM is that we treat the parameters of our planet candidates -i.e $\{\boldsymbol{x}\} = \{(R_p, \log P)\}$- as random variables, which aren't precisely known. The HBM will have 3 levels of random variables, with conditional relations between them. In our model, the data we measure - the 1st level - will depend on the true planet parameters - the second level. The detect planet parameters will themselves depend on 
%
%Inferring the parameters in an HBM is more complicated than for traditional models like \ref{eq:like

% This short bit is potentially a candidate for an Appendix

\begin{align}
L_i & = \left[ \prod_{j=1}^{N_B} \frac{(\Lambda_i (B_j))^{d_{ij}}}{d_{ij}!} \right] \exp \left(-\sum_{j=1}^{N_B} \Lambda_i (B_j) \right) \\
    & = \left[ \prod_{j=1}^{N_B} \frac{(\Lambda_i (B_j))^{d_{ij}}}{d_{ij}!} \right] \exp \left(- N_{\text{exp,i}} \right)
\label{eq:poisson-general-likelihood-nexp}
\end{align} 

where we have the expectation of the total number of planet detections for the $i^{\text{th}}$ star,  $N_{\text{exp,i}} = \sum_{j=1}^{N_B} \Lambda_i (B_j)$. Assuming that the bins $\{B_j\}_{j=1} ^{N_B}$ fully tile our parameter space, the value of $N_{\text{exp,i}}$ is independent from how we choose the bin boundaries.

Eq. \ref{eq:poisson-general-likelihood-nexp} works for an arbitrary set of bins, as long as they are disjoint, and cover the entire parameter space. Taking this to an extreme, we can choose the bin boundaries such that for each detected planet, there exists a bin which contains that planet parameters, and only its parameters; thus no bin holds two planets. We can also arbitrarily decrease the size of the bins that contain a planet. In the limit of infinitesimally small bins, we can write:

\begin{align}
\Lambda_i (B') & = \int_{B'} \lambda_i (\boldsymbol{x}) d\boldsymbol{x} \\
    & = \lambda_i (\boldsymbol{x}_{B'}) |B'|
\label{eq:small-bin-measure}
\end{align}

% This is true assuming \lambda is a continuous function; must this be stated?

where $\boldsymbol{x}_{B'}$ is a point in the bin $B'$, $|B'|$ is the volume of the bin, and we denote the new set of bins with a prime to differentiate for later. Since only the bins containing planets will appear in the product of Eq. \ref{eq:poisson-general-likelihood-nexp}, we can write the equation as:

\begin{equation}
L_i =
\begin{cases}
\displaystyle \left[ \prod_{l=1}^{d_i} \lambda_i (\boldsymbol{x}_{il}) |B_{il}'| \right] \exp \left(- N_{\text{exp,i}} \right), & \text{for } d_i \geq 1 \\[20pt]
\displaystyle \exp \left(- N_{\text{exp,i}} \right), & \text{otherwise}
\end{cases}
\label{eq:poisson-likelihood-final}
\end{equation}

where $d_i$ is the total number of planet detections for star $i$, and we use $l$ to index over the planets detected around star $i$; thus $\boldsymbol{x}_{il}$ are the parameters of the $l^{\text{th}}$ planet of star $i$.

The above still assumes that we have precise knowledge of the true planet parameters, $\boldsymbol{x}_{il}$. In reality however, the true parameters $\boldsymbol{x}_{il}$ of a detected planet are not known. Instead, our observations produce a set of measured values, $\boldsymbol{\tilde{x}}_{il}$, drawn from some uncertainty distribution, $p(\boldsymbol{\tilde{x}}_{il}|\boldsymbol{x}_{il},\boldsymbol{\alpha}_{il})$, where $\boldsymbol{\alpha}_{il}$ parametrise the uncertainty distribution. For a Gaussian uncertainty distribution, $\boldsymbol{\alpha}_{il}$ can be the variance, i.e. the error-bars squared, with $\boldsymbol{x}_{il}$ as the mean. $\boldsymbol{\alpha}_{il}$ itself depends on the lightcurve quality; the stellar parameters and their uncertainties; and so on.

The likelihood we are now interested in is the likelihood of measuring particular values for a set of detected planets. It includes two probabilities. Firstly, the probability of a set of planets being detected, given some model parameters for the detection rate, $\boldsymbol{\eta}$, and defined in terms of the planets' true parameters, $\boldsymbol{x}_{il}$. And secondly, the probability of measuring the values $\boldsymbol{\tilde{x}}_{il}$ for those parameters. We denote this new likelihood with a tilde, $\tilde{L}_i$; if the $i^{\text{th}}$ star has at least one planet detection, the likelihood is:

\begin{align}
    \tilde{L}_i & = p \left(\{\boldsymbol{\tilde{x}}_{il}\}_{l=1}^{d_i} \;\middle|\; \boldsymbol{\eta}, \{\boldsymbol{\alpha}_{il}\}_{l=1}^{d_i} \right) \\
    & = \int p \left(\{\boldsymbol{\tilde{x}}_{il}\}_{l=1}^{d_i} \;\middle|\; \{\boldsymbol{x}_{il}\}_{l=1}^{d_i}, \{\boldsymbol{\alpha}_{il}\}_{l=1}^{d_i} \right) p \left(\{\boldsymbol{x}_{il}\}_{l=1}^{d_i} \;\middle|\; \boldsymbol{\eta} \right) \textstyle\prod _{l=1}^{d_i} d\boldsymbol{x}_{il} \\
    & = \int \left[ \prod_{l=1}^{d_i} p \left(\boldsymbol{\tilde{x}}_{il} \;\middle|\; \boldsymbol{x}_{il}, \boldsymbol{\alpha}_{il} \right)\right] \left[ \prod_{l=1}^{d_i} \lambda_i (\boldsymbol{x}_{il}) |B_{il}'| \right] \exp \left(- N_{\text{exp,i}} \right) \textstyle\prod _{l=1}^{d_i} d\boldsymbol{x}_{il}
    \label{eq:halfway-derivation-uncertain-likelihood}\\
    & = \left[ \prod_{l=1}^{d_i} |B_{il}'| \int \lambda_i (\boldsymbol{x}_{il}) p \left(\boldsymbol{\tilde{x}}_{il} \;\middle|\; \boldsymbol{x}_{il}, \boldsymbol{\alpha}_{il} \right) d\boldsymbol{x}_{il} \right] \exp \left(- N_{\text{exp,i}} \right)
    \label{eq:halfway-derivation-uncertain-likelihood-2}
\end{align}

Eq. \ref{eq:halfway-derivation-uncertain-likelihood} is obtained by substituting the previously derived form of the likelihood for a Poisson point process, Eq. \ref{eq:poisson-likelihood-final} into $p \left(\{\boldsymbol{x}_{il}\}_{l=1}^{d_i} \;\middle|\; \boldsymbol{\eta} \right)$, and using the fact that the measured values of the planets are conditionally independent given $\boldsymbol{\alpha}$, which in principle includes the stellar parameters and their uncertainties.\footnote{For multiple planets orbiting a single star, the measured planet parameters are not fully independent. They are instead mutually correlated to the estimated stellar radius and mass. However, for this step we only need them to be conditionally independent, given some values for the stellar parameters.} \response{It should also be noted that this assumes that the detection efficiencies for multiple planets orbiting the same star are independent of each other.} If the $i^{\text{th}}$ star has no planet detections, there is no parameter uncertainty, and the likelihood is:

\begin{equation}
\tilde{L}_i =\displaystyle \exp \left(- N_{\text{exp,i}} \right)
\label{likelihood-no-planet}
\end{equation}

We now consider the likelihood over the entire set of stars in our sample, $\tilde{L} = \prod_{i=1}^{N_{\star}}\tilde{L}_{i}$. We can write it as a product of the individual stars' likelihoods, because we assume that the planet detections we make for individual stars are independent of each other. Most stars will not have any detected planets; the likelihood functions of these stars will come from Eq. \ref{likelihood-no-planet}. The product in Eq. \ref{eq:halfway-derivation-uncertain-likelihood-2} will only appear where there is a detected planet.  Additionally, we can ignore any multiplicative constants in our likelihood; so we drop $|B_{il}'|$. The total likelihood is then:

\begin{align}
    \tilde{L} & \propto \left[ \prod_{l=1}^{D} \int \lambda_{i(l)} (\boldsymbol{x}_{i(l)l}) p \left(\boldsymbol{\tilde{x}}_{i(l)l} \;\middle|\; \boldsymbol{x}_{i(l)l}, \boldsymbol{\alpha}_{i(l)l} \right) d\boldsymbol{x}_{i(l)l} \right] \prod_{i=1}^{N_{\star}} \exp \left(- N_{\text{exp,i}} \right) \\
    & = \left[ \prod_{l=1}^{D} \int \lambda_{i(l)} (\boldsymbol{x}_{i(l)l}) p \left(\boldsymbol{\tilde{x}}_{i(l)l} \;\middle|\; \boldsymbol{x}_{i(l)l}, \boldsymbol{\alpha}_{i(l)l} \right) d\boldsymbol{x}_{i(l)l} \right] \exp \left(- N_{\text{exp}} \right)
    \label{eq:halfway-derivation-uncertain-likelihood-3}
\end{align}

where we are now using $l$ to index over all $D$ detected planets in our entire stellar sample, and where $i(l)$ denotes the index of the star that hosts the $l^{\text{th}}$ planet. The quantity $N_{\text{exp}} = \sum_{i=1}^{N_{\star}}  N_{\text{exp,i}}$ is the expectation of the total number of detected planets in our entire stellar sample (see Sect. \ref{subsec:precise-model}).

We now reintroduce our previous parametrisation for $\lambda_i (\boldsymbol{x}_{il})$: we split the parameter space $\boldsymbol{x} = (R_p, \log P)$ into bins, within which the completeness, occurrence rate, and therefore $\lambda_i$, are constant (see Sect. \ref{subsec:precise-model}). This allows us to split the integrals in Eq. \ref{eq:halfway-derivation-uncertain-likelihood-3} into sums of integrals over each bin. The set of bins we use here, $\{B_j\}_{j=1} ^{N_B}$, is not related to the set of bins we used to obtain Eq. \ref{eq:small-bin-measure}. The likelihood becomes:

\begin{align}
    \tilde{L} & \propto \left[ \prod_{l=1}^{D}\sum_{j=1}^{N_B} \lambda_{i(l)j} \int_{B_j}  p \left(\boldsymbol{\tilde{x}}_{i(l)l} \;\middle|\; \boldsymbol{x}_{i(l)l}, \boldsymbol{\alpha}_{i(l)l} \right) d\boldsymbol{x}_{i(l)l} \right] \exp \left(- N_{\text{exp}} \right) \\
    & = \left[ \prod_{l=1}^{D} \sum_{j=1}^{N_B} \lambda_{i(l)j} F_{i(l)jl} \right] \exp \left(- N_{\text{exp}} \right) \\
    & = \left[ \prod_{l=1}^{D} \sum_{j=1}^{N_B} \comp_{i(l)j} \occr_{j} F_{i(l)jl} \right] \exp \left(- N_{\text{exp}} \right)
    \label{eq:likelihood-uncertain-final}
\end{align}

where $\lambda_{ij} = \comp_{ij} \occr_{j}$ is the (constant) detection rate within bin $B_j$, for the $i^{\text{th}}$ star, and where we define:

\begin{equation}
F_{ijl} = \int_{B_j}  p \left(\boldsymbol{\tilde{x}}_{il} \;\middle|\; \boldsymbol{x}_{il}, \boldsymbol{\alpha}_{il} \right) d\boldsymbol{x}_{il}
\end{equation}

Our final likelihood in Eq. \ref{eq:likelihood-uncertain-final} may look more complicated than Eq. \ref{eq:likelihood_precise}; in practice however, it is simple to calculate. We need to calculate the terms $F_{ijl}$ only once; it does not need to be updated as we infer the occurrence rate. We can calculate $F_{ijl}$ with a Monte Carlo sampling scheme for example. The same is true for the completeness $\comp_{ij}$, as before (see Sects. \ref{sec:detection-sensitivity} and \ref{subsec:precise-model}). The quantity $N_{\text{exp}}$ is the same as in Sect. \ref{subsec:precise-model}, and can be calculated with Eq. \ref{eq:N_exp}; it is independent of the planet detections themselves. It is also not difficult to calculate the Jacobian of $\tilde{L}$ w.r.t $\boldsymbol{\occr}$, which allows us to later use gradient-based sampling methods (see Sect. \ref{sec:posterior}).

The method presented here is general, and can be applied to any occurrence rate study where the completeness and occurrence rate distribution are parametrised with a piecewise-constant step-function, which is very popular in the literature \citep[see e.g][]{howard2012,dong2013,swift2013,dressing2015,mulders2015a,he2017}. The alternative would have been to infer the full joint-posterior distribution of the HBM, including a set of parameters $\boldsymbol{x} = (R_p, \log P)$ for each detected planet. This would add 2 model parameters per detected planet, and more if we wanted to find the relation between planetary occurrence rates and other parameters, such as stellar metallicity or stellar mass. With the single planet detection in our study, the difference is not large. In general, however, studies such as \cite{dressing2015} can include hundreds of planet candidates, which rapidly inflates the number of parameters we need to infer. Directly marginalising over the planet parameters can therefore lead to significant increase in computational speed with such large planet samples, which our method avoids.

\subsection{Posterior distribution of $\mathbf{\eta}$} \label{sec:posterior}

We infer the occurrence rates within a Bayesian framework, and our posterior distribution on $\boldsymbol{\occr}$ is given by:

\begin{equation}
    p(\boldsymbol{\occr}|\text{data}) \propto \tilde{L}(\boldsymbol{\occr}) \pi(\boldsymbol{\occr})
\end{equation}

where we use the Jeffreys prior for the rate parameters of a Poisson process:

\begin{equation}
    \pi(\boldsymbol{\occr}) = \prod_{j=1}^{N_B} \frac{1}{\sqrt{\occr_j}}
\end{equation}

% I don't mention here that I'm actually drawing samples frmo log(occr) as that converges much quicker (samples occr takes an order of magnitude longer to converge), but since it's equivalent and just a computational thing, better not to waste space.

We draw samples from the posterior using Markov Chain Monte Carlo sampling methods. When dealing with a smaller number of model parameters (in Sect. \ref{sec:occr-constrained}), we use the affine-invariant sampler implemented in \texttt{emcee} \citep[see][]{goodman2010,dfm2013}; however, for larger numbers of parameters we find that it doesn't perform as well as gradient-based samplers. As we have closed-form equations for our likelihood and our prior, we also directly calculate the gradient of $p(\boldsymbol{\occr}|\text{data})$ w.r.t $\boldsymbol{\occr}$. This allows us to use Hamiltonian Monte Carlo samplers, such as the No-U-Turn Sampler implemented in \texttt{PyMC3}, for much faster convergence \citep{hoffman2014,salvatier2016_pymc3}.

%\subsubsection*{Model variants}

%%%%%%%%%%%%%%%%%%%%%%%%%%%%%%%%%%%%%%%%%%%%%%%%%%%%%%%%%%%%%%%%%%%%%%%%%%%%%%%%

\section{Results} \label{sec:results}

% This is where I mention that we didn't find anything I guess.
% This is a space for just the dry results.
% Plot occurrence rates integrated over full period (to 20 days), and integrated over smaller prior (to 5 days).
% Other papers (Demory 2016) only look at the innermost planets, i.e b and c, so the comparison to them should also cover a similar period range.

\begin{figure*}
    \centering
    \includegraphics[scale=1.0]{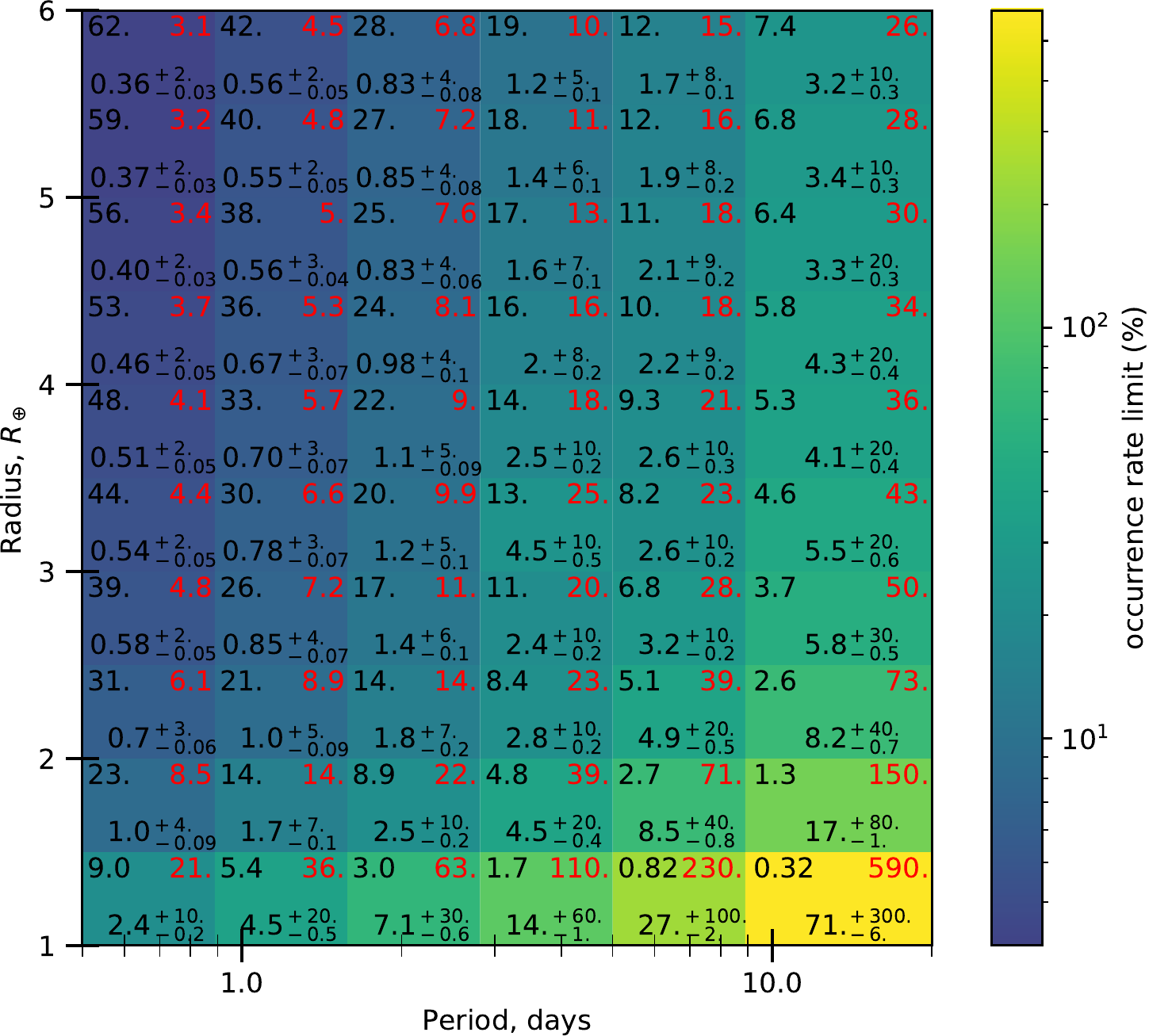}
    \caption{The inferred occurrence rate - in \% - in our sample as a function of $(R_p, \log P)$. Within each bin, the numbers refer to the following quantities: \textit{top right:} upper bound in 95\% credible interval, i.e the 95-th percentile; \textit{top left:} the expected number of detections given an occurrence rate $\occr=1$ within the bin, equal to $\frac{1}{\tr \ds}$; \textit{bottom:} the median and $84\%$ credible intervals of the occurrence rate.}
    \label{fig:occr-grid}
\end{figure*}

\begin{figure}
    \centering
    \includegraphics[width=\hsize]{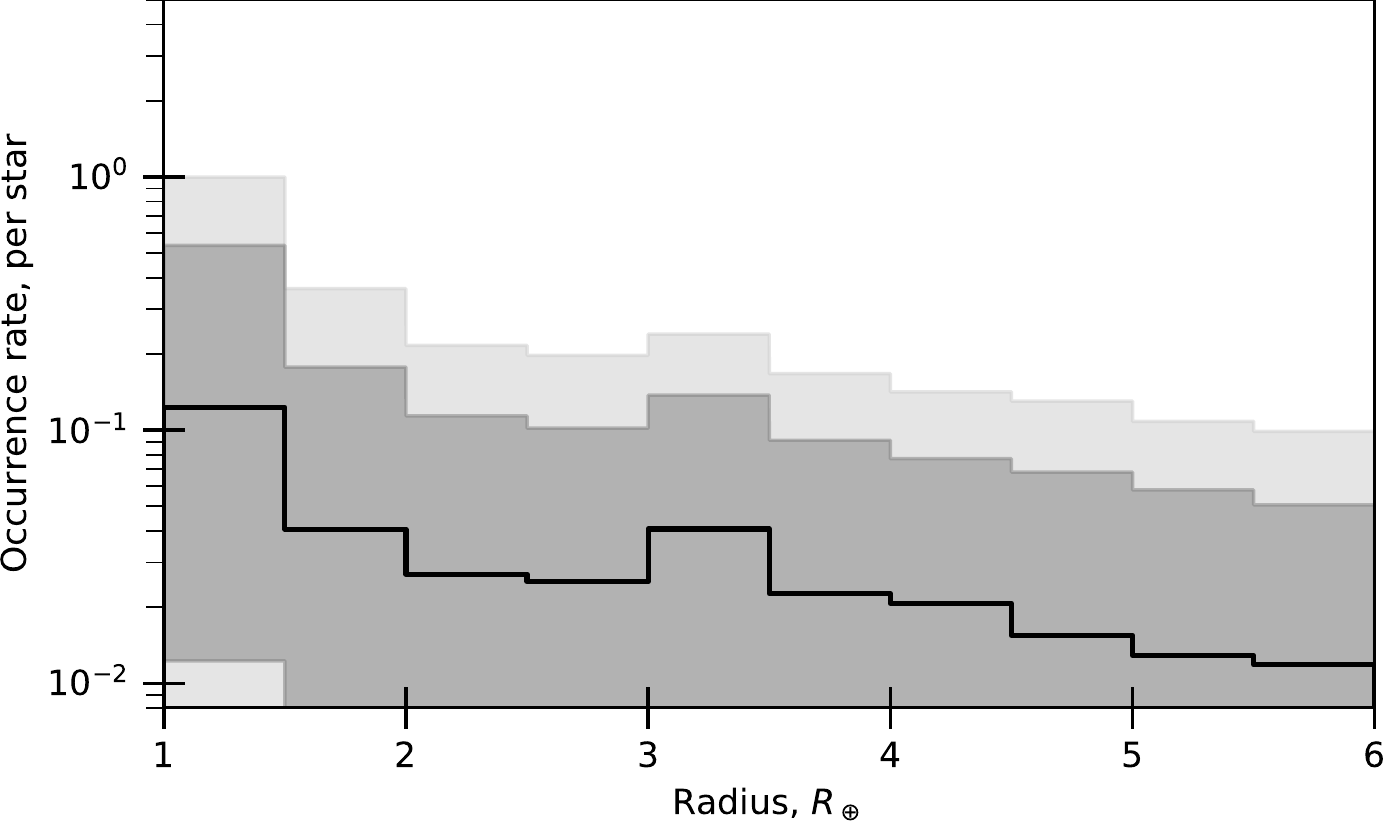}
    \caption{The occurrence rate in our sample as a function of planet size. It is integrated over orbital periods of \res{1.0-20} days, assuming a uniform occurrence rate in $\log P$ within that range. The grey shaded regions show $84\%$ and $95\%$ credible intervals.}
    \label{fig:occr-period}
\end{figure}

\subsection{Occurrence rates} \label{sec:occr-unconstrained}

% TODO: update the numbers if anything changes.

 \begin{table}
 \caption{Occurrence rates for different planet classes: upper bounds on the 95\% credible interval preceded by $<$ where no detections are found, and median and $84\%$ credible intervals as uncertainties for the bin containing K2-25b. This is for the case where we do not constrain the occurrence rate to be uniform in $\log P$.}
 \centering{}%
 \begin{tabular}{l  c c c }
 \hline 
 Period range & & Radius range, $\si{\Re}$ & \tabularnewline
 days & $1 - 2$ & $2 - 4$ & $4 - 6$\tabularnewline
 \hline 
 $0.5 - 0.9$ & $<0.24$ & $<0.12$ & $<0.08$\tabularnewline
 $0.9 - 1.6$ & $<0.41$ & $<0.17$ & $<0.12$\tabularnewline
 $1.6 - 2.8$ & $<0.70$ & $<0.28$ & $<0.18$\tabularnewline
 $2.8 - 5.0$ & $<1.29$ & $0.21^{+0.17}_{-0.11}$ & $<0.31$\tabularnewline
 $5.0 - 8.9$ & $<2.58$ & $<0.69$ & $<0.42$\tabularnewline
 $8.9 - 20.0$ & $<6.40$ & $<1.28$ & $<0.73$\tabularnewline
 %\hline
 %$0.5 - 5.0$ & $<1.88$ & $<0.83$ & $<0.48$ \tabularnewline
 \hline
 \end{tabular} \\
 \label{tab:occr-table}
 \end{table}
 
We consider radii between $1 \si{\Re}$ and $6 \si{\Re}$, split into bins with a $0.5\si{\Re}$ interval; and periods between $0.5$ and $20$ days, split into regular intervals in log-period. Given only one detected planet in the sample, our constraints on the occurrence rate act mainly as upper limits. \response{When referring to the upper limit on the occurrence rate, we will henceforth refer to 95-th percentile in the posterior distributions for $\occr$; i.e. the upper bound on the 95\% credible interval}. Shown in red in Fig. \ref{fig:occr-grid}, the occurrence rate upper limits follow a roughly inverted pattern from the completeness in \ref{fig:completeness}. The upper-left numbers in Fig. \ref{fig:occr-grid} show the expected number of planets we would detect, if the occurrence rate had been 1 within the bin. At larger radii and lower periods, where we have a better completeness, our constraints are stronger and lower, down to a few percent per bin. At radii below $1.5\si{\Re}$ on the other hand, the occurrence rate cannot be constrained to below $1.0$ for longer orbital periods, due to the low completeness of our sample.

% Include things like "number of planets detected in total if the occurrence rate was 1" for context.

% TODO: Does the new language (95% credible intervals are below...) make sense? Or perhaps, we place upper bounds on the ... Ask Brice specifically.

For super-Earths ($1-2\si{\Re}$), \response{we find the occurrence rate to be in 95\% credible intervals below} \res{0.24, 0.41, 0.70, 1.29}, for period ranges of $0.5-0.9$, $0.9-1.6$, $1.6-2.8$ and $2.8-5.0$ days, respectively. Our constraint on the mini-Neptune ($2-4\si{\Re}$) population is stronger. \response{Within the same period ranges, the 95\% credible intervals for mini-Neptunes are below} \res{0.12, 0.17, 0.28, 0.53}. In the $2.8-5.0$ day bin, where we detect the sub-Neptune K2-25b, we also find an occurrence rate of \res{$0.21_{-0.11}^{+0.17}$} for mini-Neptunes. For planets above the size of Neptune ($4-6\si{\Re}$, referred to as gas giants from here), we find that there are fewer than \res{0.08, 0.12, 0.18, 0.31} per star within those period ranges. The results are summarised in Table \ref{tab:occr-table}. Note that the 95\% upper limit integrated across multiple bins is not equal to the sum of the 95\% upper limits within each bin, as the upper limit depends on the distribution variance, which is not additive.

\subsection{Occurrence rates integrated over period} \label{sec:occr-constrained}

\begin{center}
\begin{table}
\caption{Summary for the inferred distributions on the occurrence rates of super-Earths, mini-Neptunes, and ice/gas giants. The occurrence rate in this case is constrained to be uniform in $\log P$ for 1-20 days.}
\centering{}%
\begin{tabular}{l  c c c }
\hline 
 & & Radius range, $\si{\Re}$ & \tabularnewline
 & $1.00 - 2.00$ & $2.00 - 4.00$ & $4.00 - 6.00$\tabularnewline
\hline 
Estimate & $--$ & $0.20^{+0.16}_{-0.11}$ & $--$\tabularnewline
$84\%$ upper limit & $0.66$ & $0.36$ & $0.21$\tabularnewline
$95\%$ upper limit & $1.14$ & $0.49$ & $0.29$\tabularnewline
\hline
\end{tabular} \\
\label{tab:occr-classes}
\end{table}
\par\end{center}

To estimate the total occurrence rate within a period range such as $1-20$ days, we must sum the rate estimates over that range at a sample level. If we do this, however, the result will be dominated by the most poorly-constrained bins; in other words, the bins at the upper end of the period range, where the completeness is lowest.

We can instead make the assumption that the occurrence rate is uniform in the log-period, within $1-20$ days. This is consistent with previous results for earlier stellar classes \citep[see e.g][]{mulders2015a,mulders2018}, and with the periods of the TRAPPIST-1 system. We re-run our analysis under this constraint, and integrate the occurrence rates within $1-20$ days for each radius bin. The results are shown in Fig. \ref{fig:occr-period}. This is equivalent to calculating the occurrence rates with one large bin covering our entire period range.

The distributions on the resulting occurrence rates for super-Earths ($1-2\si{\Re}$), mini-Neptunes ($2-4\si{\Re}$) and ice/gas giants ($4-6\si{\Re}$) are summarised in Table \ref{tab:occr-classes}. For super-Earths and ice/gas giants, where we have no detections, \res{we find 95\% upper limits of $1.14$ and $0.29$} planets per star. For mini-Neptunes, with one detection, \res{we find an occurrence rate of $0.20^{+0.16}_{-0.11}$ planets per star, with a 95\% upper limit of 0.54}.

\subsection{TRAPPIST-1-like systems}
\label{sec:trappist-results}

% Not anymore: I refer to this section several times for the following point: the chance of finding the TRAPPIST-1 system is essentially equal to the chance of finding TRAPPIST-1b. Therefore, it assumes that the inner planet of such a system has period 1.5 days.

We do not find any planetary systems like TRAPPIST-1 in our sample. We can take our completeness for TRAPPIST-1-like systems to be the probability of finding at least one of the planets, if all seven were orbiting a star in our sample. With low mutual inclination, the completeness will depend on our ability to detect the most-detectable planet in the system, TRAPPIST-1b. If we assume an occurrence rate for TRAPPIST-1-like systems within our sample, we can then calculate the probability of not finding any in our data (or conversely the probability of finding at least one). In the case where every ultracool dwarf in our sample held a system such as TRAPPIST-1, we would have a \res{4.2\%} chance of finding none (or a \res{95.8\%} chance of finding at least one). For a 20\% and 5\% system occurrence rate, the probability of finding at least one would be \res{46\%} and \res{14\%}, respectively.

%%%%%%%%%%%%%%%%%%%%%%%%%%%%%%%%%%%%%%%%%%%%%%%%%%%%%%%%%%%%%%%%%%%%%%%%%%%%%%%%

\section{Discussion} \label{sec:discuss}

% \subsection{Trends in occurrence rates}

% Mini-Neptunes are interesting because they contain 95% of the heavy element mass of M Dwarf planetary systems (Mulders 2015, increase in mass...)
% Reference Mordasini 2018

We find that gas giant planets are uncommon around ultracool stars, with fewer than $0.28$ per star. We also improve upon the \citet{demory2016} and \citet{he2017} constraints of mini-Neptune and super-Earth populations. \response{Our results on the 95\% upper limits are also in agreement with \cite{sagear2019} across all planet classes.}

The rarity of gas giants around low-mass stars has already been discussed in the literature \cite[see e.g.][]{santos2004,johnson2010,gaidos2013_1,obermeier2016,mulders2018}. Short-period sub-Neptunes however, have been found to become more common as stellar mass decreases \citep[see e.g.][]{howard2012,dressing2015,mulders2015a,mulders2015b,muirhead2015}. Collating the results of several studies in a review, \citet{mulders2018} showed that sub-Neptunes become $\sim3$ times more common around M dwarfs, relative to FGK stars.\footnote{Here we differentiate between sub-Neptunes - meaning all planets smaller than Neptune, including super-Earths and rocky planets - and mini-Neptunes, which we define as having radii $2-4\si{\Re}$ in this paper.} The increased prevalence of small planets around low-mass stars is also predicted by formation synthesis models \citep[e.g.][]{alibert2017,liu2019}. \citet{alibert2017} find however, that larger planets such as mini-Neptunes are not reproduced by formation models for ultracool dwarfs. Whether the trend of increasing small planet occurrence rates continues into the late-M and brown dwarfs has thus far not been observationally determined.

% \todo{Cite other occurrence rate papers here, MASSIVE, etc... Also cite theoretical papers: Remo when it comes up, etc...}.

% In a review of the literature, \citet{mulders2018} collates the results of several studies on earlier M and late-K dwarfs, assuming a uniform occurrence rate in log-period and log-radius. They find that at $P<50$ days and radii of $1-4 \si{\Re}$, the occurrence rate is around 1.5 planets per star. Whether this trend continues into late-M dwarfs and brown dwarfs has been an open question.}

As an illustration of this trend, \citet{fressin2013} found $0.066\pm0.004$ mini-Neptunes ($2-4\si{\Re}$) and $0.096\pm0.001$ super-Earths ($1.25-2\si{\Re}$) per star, at periods of $0.8-17\si{d}$ for FGK stars in \textit{Kepler}. In the sample of mostly mid- and early-M dwarfs observed by \textit{Kepler} on the other hand, \citet{dressing2015} found $0.29$ planets with radii $2-4 \si{\Re}$ and periods $0.9 < P < 18.3\si{d}$, and $0.57$ with radii $1-2 \si{\Re}$ (these are not upper limits, but averages). The other studies collated by \citet{mulders2018} show similar results, see e.g \citet{youdin2011,howard2012,dong2013,petigura2013,morton2014,mulders2015b,silburt2015,gaidos2016}, and also \cite{hsu2019,hsu2020}.

Assuming uniform occurrence rates with log-period as in \citet{mulders2018} - without assuming uniformity in log-radius - we find that there are \res{$0.20^{+0.16}_{-0.11}$} mini-Neptunes per ultracool star (see Sect. \ref{sec:occr-constrained}). This is in line with the results for earlier M dwarfs, showing that the trend of high mini-Neptune occurrence rates is maintained for ultracool dwarfs. Such mini-Neptunes are larger than the range of radii reproduced by \citet{alibert2017}, though this result relies on the detection of the mini-Neptune K2-25b, which appears to have an unusually large planet/star mass ratio.\footnote{Using the empirical M-R relation derived by \citet{wolfgang2016}, K2-25b would have a mass of $13.4\pm1.9\si{\Me}$, which leads to a planet/star mass-ratio of \res{$2\times10^{-4}$} \citep[with the stellar mass estimate from][]{mann2016}, larger than the $3\times10^{-5}$ cutoff found by \citet{pascucci2018}. \cite{mann2016} note that it's possible that K2-25b's large radius could be due to its relatively young age, and that it may not have had the time to lose its hydrogen envelope.}

%, and their masses may be above the predicted pebble isolation cutoff \citep{liu2019} or the previously-found threshold of 

Extending the comparison to super-Earths ($1-2\si{\Re}$), we find that the integrated occurrence rate within $1-20d$ is \response{within a 95\% credible interval} below \res{1.14}, and \response{within a 84\% credible interval} below \res{0.66}. \response{Using a well-characterised sample of M-dwarfs, \cite{hardegree-ullman2019} found that they host $0.27^{0.25}_{-0.18}$ planets per star, at periods of $0.5-10$ days and radii of $1.5-2.5\si{\Re}$. To compare, our results within $1-10$ orbital periods place a 95\% upper limit of fewer than $0.36$ planets of such size per star. The comparison depends on the assumed distribution of underlying planetary orbital periods: we assume that the occurrence rate density is uniform in log-period, while \cite{hardegree-ullman2019} set their distribution in period by sampling from the observed periods of their sample of 13 detected planets, 12 of which follow a roughly uniform distribution between 1-10 day periods. \cite{muirhead2015} similarly find that $21^{+7}_{-5}\%$ of mid-M dwarfs host compact multi-planet systems with orbital periods less than 10 days.}

\response{Overall, when compared with previous studies on mid- and early-M dwarfs, our results do not exclude a continuing trend of increasing occurrence rates for sub-Neptunes around lower mass stars. Further comparison would require more lower limits on the occurrence rate, which will be possible with more planetary detections.}

With our completeness, if every ultracool dwarf hosted a TRAPPIST-1-like system, we would have a \res{$\sim96\%$} chance of finding at least one, based on the detectability of TRAPPIST-1b. This is strongly dependent on the $1.5$ day period of the inner planet however. Based on these facts, and that we don't detect any such systems, we can conclude that such short-period systems are not ubiquitous. The inner-planet periods may however be longer than those in TRAPPIST-1, allowing them to remain undetected in our sample. It is also possible that the formation of mini-Neptunes around $\sim20\%$ of stars may inhibit the formation of long chains of rocky planets.

It has been suggested that inward migration of solids can explain the increased occurrence rate of short-period sub-Neptunes around low-mass stars \citep[see e.g.][]{mulders2015a}. Otherwise, we would expect that the lower disk masses M dwarfs should produce fewer and/or lower-mass planets. In future studies, it will be informative to test this hypothesis against spectral type/stellar mass, by directly constraining the distribution of occurrence rates as function of stellar mass. Instead of treating a sample of stars as a homogeneous set, an occurrence rate model could include a functional relation between the occurrence rate and stellar mass. With such data, we could directly test formation predictions that depend on stellar mass, such as pebble isolation enforcing a maximum super-Earth mass \citep{liu2019}. However, this would make the computation more expensive, since simplifications such as Equation \ref{eq:likelihood_star_product} into \ref{eq:likelihood_precise} would no longer be possible.

Given that TESS will not be sensitive to TRAPPIST-1-like systems around ultracool dwarfs (Sebastian at al. 2020, \textit{in prep}), and \textit{K2} has finished its mission, our main chance of finding more rocky planets for atmospheric characterisation with JWST will come from ground-based surveys such as SPECULOOS and SAINT-EX. These surveys are aimed specifically at finding earth-sized planets around ultracool dwarfs, and they will provide much-needed completeness at low radii \citep{delrez2018}. The downside will be that they have shorter observation timescales, and less completeness at longer orbital periods. Moreover, modelling their completeness isn't trivial due to the daily gaps in the data, and the likelihood of having to use single-transits as a detection criterion.

While most occurrence rate studies tend to use data from a single survey, it is also possible to use the model in Sect. \ref{sec:model} on multiple surveys, with different completeness functions. Combining the initial results of the SPECULOOS and SAINT-EX surveys with those of \textit{K2} will be crucial in further constraining the occurrence rate of earth-sized planets. If such planetary systems prove to be uncommon, TRAPPIST-1 is likely to remain one of the unique few rocky planet systems that can be characterised by JWST.

%%%%%%%%%%%%%%%%%%%%%%%%%%%%%%%%%%%%%%%%%%%%%%%%%%%%%%%%%%%%%%%%%%%%%%%%%%%%%%%%

%\section{Conclusion} \label{sec:conclusion}

%%%%%%%%%%%%%%%%%%%%%%%%%%%%%%%%%%%%%%%%%%%%%%%%%%%%%%%%%%%%%%%%%%%%%%%%%%%%%%%%

\begin{acknowledgements}
The authors would like to thank Brett Morris for his helpful comments.

M.S acknowledges support from the Swiss National Science Foundation (PP00P2-163967). B.-O.D. acknowledges support from the Swiss National Science Foundation in the form of a SNSF Professorship (PP00P2-163967). This work has been carried out within the framework of the NCCR PlanetS supported by the Swiss National Science Foundation. Calculations were performed on UBELIX (http://www.id.unibe.ch/hpc), the HPC cluster at the University of Bern. \\

This publication makes use of VOSA, developed under the Spanish Virtual Observatory project supported by the Spanish MINECO through grant AyA2017-84089. VOSA has been partially updated by using funding from the European Union's Horizon 2020 Research and Innovation Programme, under Grant Agreement nº 776403 (EXOPLANETS-A).\\

This paper includes data collected by the Kepler mission and obtained from the MAST data archive at the Space Telescope Science Institute (STScI). Funding for the Kepler mission is provided by the NASA Science Mission Directorate. STScI is operated by the Association of Universities for Research in Astronomy, Inc., under NASA contract NAS 5–26555. This study made use of data obtained within GO programs 2010, 2011, 3011, 3044, 3045, 3114, 4026, 4030, 4039, 4036, 4081, 5026, 5030, 5036, 5039, 5081, 6002, 6005, 6006, 7002, 7005, 7006, 8030, 8054, 10030, 10054, 11004, 11029, 11042, 12004, 12029, 12042, 13004, 13029, 13042, 14018, 14067, 15018, 15067, 16018, 16067, 17008, 17057, 18008, 18057 and 19008.
\end{acknowledgements}

% Bibliography
% ------------

\bibliographystyle{aa} % style aa.bst
\bibliography{occr_background,k2gp_technical,formation_background,stellar_catalogues_characterisation,additional_references} % your references Yourfile.bib
%
% - join the .bib files when you upload your source files
%------------------------------------------------------------------

\end{document}